\documentclass{aa} %manuscript 
\usepackage{graphicx}
\usepackage{txfonts}
\usepackage{hyperref}

\hypersetup{
    colorlinks=true,
    linkcolor=blue,
    urlcolor=blue,
    citecolor = blue,
}

\begin{document} 

   \title{Caveats about measuring carbon abundances in stars using the CH band} %, even at high spectral resolution} %( using the X-shooter Spectral Library (XSL)}
   
   \author{P. Santos-Peral
          \inst{1,2}
          \and
          P. S\'{a}nchez-Bl\'{a}zquez\inst{1,2}
          \and
          A. Vazdekis\inst{3,4}
          \and
          P. A. Palicio\inst{5}
          \and
          A. T. Knowles\inst{3,4,6}
          \and
          A. Recio-Blanco\inst{5}
          %\and
          %I. A. Gaspar-Gorostieta\inst{3,4}
          \and
          C. Allende Prieto\inst{3,4}
          }

\institute{Departamento de F\'{i}sica de la Tierra y Astrof\'{i}sica, Universidad Complutense de Madrid, 28040 Madrid, Spain\\
              \email{pasant02@ucm.es}
        \and
             Instituto de F\'{i}sica de Part\'{i}culas y del Cosmos IPARCOS, Facultad de CC F\'{i}sicas, Universidad Complutense de Madrid, 28040 Madrid, Spain
         \and
             Instituto de Astrof\'{i}sica de Canarias, E-38200 La Laguna, Tenerife, Spain
        \and
             Departamento de Astrof\'{i}sica, Universidad de La Laguna, E-38205, Tenerife, Spain
        \and
            Université Côte d’Azur, Observatoire de la Côte d’Azur, CNRS, Laboratoire Lagrange, France 
        \and
            Data Science Research Centre, Liverpool John Moores University, Liverpool, L3 3AF, UK \\
             }

   \date{Received May 26 2025 / Accepted July 14 2025}

% \abstract{}{}{}{}{} 
% 5 {} token are mandatory
 
  \abstract
  % context heading (optional)
  % {} leave it empty if necessary  
   {Stellar carbon abundances are crucial for tracing the star formation history and predicting the near-infrared emission of galaxies. However, deriving accurate carbon abundance estimates for a wide variety of stars is still complex due to the difficulties in properly measuring it from atomic and molecular lines. Therefore, its inclusion in stellar population models remains challenging.}
  % aims heading (mandatory)
   {The aim of this paper is to analyse the carbon abundance determination for the large empirical X-shooter Spectral Library~(XSL), commonly used as a benchmark for the development of stellar population models.}
  % methods heading (mandatory)
   {The carbon abundance analysis was performed over strong molecular CH bands in the G-band region. We used the GAUGUIN automated spectrum synthesis code, and adopted two different grids of reference synthetic spectra separately, each with the same [C/Fe] abundance coverage. We carried out a detailed comparison between both grids to evaluate the accuracy and the model dependence of the measured [C/Fe] abundances.}
  % results heading (mandatory)
   {We obtained a large and precise unbiased [C/Fe] abundance catalogue ($\sim$~200 stars) from both theoretical grids, well distributed in the Hertzsprung-Russell (HR) diagram and with no trend with the stellar parameters. We also measured compatible values from each independent CH band, with a high-quality [C/Fe] abundance estimate for both dwarfs and giants indistinctly. We always observed a dispersed flat trend around [C/Fe]~$\sim$~0.0~dex all along the covered metallicity regime (-2.5~$<$~[Fe/H]~$<$~+0.5~dex), in agreement with some literature studies. However, we reported variations up to $\arrowvert\Delta[\rm C/Fe]\arrowvert$~$\sim$~0.8~dex in the [C/Fe] composition of the star depending on the adopted grid. We did not find such differences in the $\alpha$-element measurements. This behaviour implies a strong model dependence in the [C/Fe] abundance estimate. }
  % conclusions heading (optional), leave it empty if necessary 
   {Potential sources of error could be associated with the use of spectral synthesis methods to derive stellar carbon abundances in the CH~4300~$\AA$ band. Intrinsic small differences in the synthetic models over this crowded and blended region may induce a large disparity in the precise abundance estimate for any stellar type, leading to inaccurate carbon measurements without being noticed.}

   \keywords{stars:abundances --
                methods: data analysis --
                techniques: spectroscopic
               }

    \maketitle
%
%________________________________________________________________

\section{Introduction}

\fancypagestyle{firstpage}
{
    \fancyhead{} 
    \fancyfoot[L]{IPARCOS-UCM-25-041}
}

The chemical composition of stars is intimately linked to the Star Formation History (SFH) of galaxies and star clusters~\citep[e.g.][]{freeman2002, harris2009, delaRosa2011, emma2018}. %that is to say, the evolution of the star formation rate over time.  
While individual stellar abundances can be directly measured in the Milky Way and nearby galaxies, the properties of more distant systems must be inferred indirectly through stellar population synthesis models~\citep[e.g.][]{BruzualCharlot2003, Molla2009, Vazdekis2010, Maraston2020}. \par

These models often rely on empirical stellar libraries that reflect the chemical abundance patterns of the Milky Way~\citep[e.g.][]{Vazdekis1999, Vazdekis2010, Vazdekis2016, BruzualCharlot2003, LeBorgne2004, Maraston2011, Maraston2020, Verro2022_models}, and the predictions, therefore, are linked to the specific star formation history of the Solar neighbourhood. However, to accurately interpret the integrated light from external galaxies, models need to make predictions for well-defined abundance patterns that do not change with metallicity. A common approach to make these predictions is to apply theoretical corrections to the empirical spectra ~\citep[e.g.][]{Cervantes2007, Walcher2009, Conroy2012, conroy2018, Vazdekis2015, LaBarbera2016, Knowles2023} using the changes in the model atmosphere spectra when a single chemical element is changed. However, this approach requires to know the chemical abundances of the empirical stars in the first place. \par

So far, carbon (C; Z~=~6) has not been fully included in stellar population models due to the lack of large empirical spectral libraries with reliable [C/Fe] measurements across a wide range of stellar parameters (T$_{\rm eff}$, log(g), [Fe/H]), such as MILES~\citep{SanchezBlazquez2006}, MEGASTAR~\citep{Carrasco2021}, ELODIE \citep{prugniel2001} or X-shooter \citep{XshooterDR1, Xshooter_Param, XshooterDR2, XshooterDR3}. Yet, carbon\footnote{For simplicity, throughout the paper, the term "carbon abundance" will always refer to the abundance of carbon relative to iron~([C/Fe]).} is essential for understanding both the early chemical enrichment of galaxies —through contributions from massive stars and Type II supernovae— and later stages of stellar evolution, particularly by asymptotic giant branch (AGB) stars and binary interaction mechanisms \citep[see e.g.][]{frebel2007, kobayashi2020}. \par

Carbon-rich stars are found at all metallicities and evolutionary stages, with a rising incidence at low metallicities, being an important imprint of early star formation episodes \citep[see e.g.][]{lebzelter2018, Sestito2024, Arentsen2025}. AGB stars (giants of low-to-intermediate masses: 1-8~M$_\odot$), in particular, dominate the near-infrared (NIR) emission in intermediate-age stellar populations, but theoretical predictions of their contribution remain uncertain~\citep[e.g.][]{Baldwin2018, DahmerHahn2018, Verro2022_models} due to the complex physics of mixing~\citep[\textit{dredge-ups}, e.g.][]{Iben1965, gallino1998}, binarity, and mass loss~\citep[e.g.][]{cristallo2011, karakas2014, osborn2025}. \par 

Additionally, there are persistent issues in the NIR stellar population models. For instance, recent models are unable to reproduce the CO-strong feature observed in the spectra of early-type galaxies~\citep[see][and references therein]{elham2022}. Moreover, there are still discrepancies between spectroscopic ages and Color-Magnitude Diagram (CMD) derived ages \citep[so-called model zero-point problem, see e.g.][]{elham2025}. These issues further underscore the need for accurate carbon abundance data in empirical libraries. \par

Observationally, determining [C/Fe] remains challenging. Reported abundance values widely vary depending on the spectral resolution, methodology, and the specific analysed carbon features. Most large surveys focus on FGK dwarfs (3500~$\lesssim$~T$_{\rm eff}$~$\lesssim$~7500~K; log(g)~$\textgreater$~3.5~cm~s$^{-2}$) because their atmospheres are less affected by internal mixing than those of giant (log(g)~$\textless$~3.5~cm~s$^{-2}$) stars \citep[e.g.][]{DelgadoMena2010, DelgadoMena2021, nissen2014, franchini2020, bensby2021}. \par

The main goal of this work is to derive reliable [C/Fe] abundances for stars in the X-shooter Spectral Library \citep[hereafter XSL;][]{Xshooter_Param, XshooterDR2}, one of the most comprehensive empirical stellar libraries currently available with enough resolution to measure individual chemical abundances. We perform a detailed spectroscopic analysis of the CH molecular bands in the G-band region for a subsample of 611 stars from XSL~\citep{SantosPeral2023}. The structure of this paper is as follows: in Sect.~\ref{data} we describe the observational data; Sect.~\ref{method} details the spectral synthesis methodology and line selection; Sect.~\ref{results} presents the [C/Fe] measurements and their uncertainties; and Sect.~\ref{conclusions} summarizes our findings.

%__________________________________________________________________

\section{The X-shooter observational data sample} \label{data}

The XSL\footnote{\url{http://xsl.astro.unistra.fr}} \citep{XshooterDR1, XshooterDR2, XshooterDR3} is an empirical stellar spectral library of a total of 830 spectra for 683 stars, and can be considered as a reference stellar library for the optical and the NIR region. The XSL sample comprises a wide variety of stellar types (3000~$\lesssim$~T$_{\rm eff}$~$\lesssim$~20000~K, 0~$\lesssim$~log(g)~$\lesssim$~+5.0~cm~s$^{-2}$, -2.5~$\lesssim$~[Fe/H]~$\lesssim$~+0.5~dex) that satisfactorily covers the HR diagram in-depth \citep[see Fig.~2 or Fig.~1 in][respectively]{Xshooter_Param, XshooterDR3}. The employed X-shooter three-arm spectrograph at the ESO's VLT telescope \citep{Vernet2011} 
has sufficient spectral resolution to measure carbon abundances in three different spectral ranges: UVB (300-556~nm; R$\sim$10000), VIS (533-1020~nm; R$\sim$11000), and NIR (994-2480~nm; R$\sim$8000). \par

However, only stellar atmospheric parameters (T$_{\rm eff}$, log(g), [Fe/H]) from the UVB and VIS arms of the X-shooter Data Release~2 \citep[DR2,][]{XshooterDR2} could be estimated by \citet{Xshooter_Param} \citep[see][for further details of the additional XSL~DR2 stellar atmospheric parameters validation with recent published datasets]{SantosPeral2023}, which led to a total of 748 spectra, corresponding to 611~stars. For the present work, we specifically focused on the UVB arm (typical S/N~$\approx$~70) due to the carbon lines selection in the G-band region around 4300~$\AA$, which corresponds to a final sample of 504 stars (later described in detail in Sect.~\ref{method_lines}~and~\ref{flags_and_errors}).

%__________________________________________________________________

\section{Method} \label{method}

To estimate [C/Fe] abundances from different spectral regions, we made use of the spectrum synthesis code GAUGUIN \citep{Bijaoui2012, guiglion2016, alejandra2016, alejandra2022_RVS} over the observed XSL data sample. \par 

GAUGUIN is an automated abundance estimation code, based on a local optimisation method around a given set of parameters (T$_{\rm eff}$, log(g), [Fe/H], [$\alpha$/Fe], [X/Fe]) linked to a reference synthetic spectrum.  This code has been developed in the framework of the Gaia-ESO Survey \citep{alejandra2014} and the GSP-Spec module for the \emph{Gaia} Astrophysical parameters inference system (Apsis) pipeline \citep[][]{alejandra2016, alejandra2022_RVS, Apsis2023}. We refer the reader to \citet{SantosPeral2020} and \citet{alejandra2022_RVS} for a complete description of the method. \par 

In this work, the four input parameters of each observed X-shooter spectrum were adopted from \citet[][T$_{\rm eff}$, log(g), {[}Fe/H{]}]{Xshooter_Param} and \citet[][for {[}$\alpha$/Fe{]}\footnote{We measured [Mg/Fe] and [Ca/Fe] abundances, its average is used as a proxy for the [$\alpha$/Fe] stellar composition.}]{SantosPeral2023}.

\subsection{Reference synthetic spectra grids} \label{method_grids}

We used two different computed sets of 5D reference synthetic spectra grids, varying in the T$_{\rm eff}$, log(g), [M/H], [$\alpha$/Fe], and [C/Fe] dimension: a theoretical library used for the computation of the semi-empirical MILES stellar spectra \citep[sMILES;][]{Adam2021}, and the updated BOSZ synthetic stellar spectral library \citep{Bosz2024}. We performed a detailed comparison between both grids in order to draw more robust conclusions about the accuracy and the model dependence of the derived [C/Fe] abundances.

\subsubsection{Theoretical library used to compute sMILES: semi-empirical MILES stellar spectra \citep{Adam2021}} \label{method_Adam}

sMILES is a library of semi-empirical stellar spectra that is based on the empirical Medium resolution Isaac Newton Library of Empirical Spectra \citep[MILES;][]{SanchezBlazquez2006} and the differential abundance pattern predictions of theoretical spectra. A new high-resolution (R~$\sim$~100000) theoretical stellar spectral library\footnote{\url{https://uclandata.uclan.ac.uk/178/}} was specifically computed for that project (wavelength coverage: $\Delta\lambda$~=~1677–9001~$\AA$,  with a wavelength step of 0.05~$\AA$), which has also been used in the present analysis. \par

This grid of theoretical spectra was generated using \texttt{ATLAS9} 1D LTE model atmospheres \citep{Kurucz1979,kurucz1993}, including variations in metallicity, $\alpha$-elements enhanced in lock-step (O, Ne, Mg, Si, S, Ca, and Ti) and carbon abundances. The spectra were computed using the ASSET radiative transfer code \citep{Koesterke2009}, based on \texttt{SYNSPEC} routines from \citet{Hubeny2000, Hubeny2017}. They adopted the solar chemical composition of \citet[][log~$\epsilon_{\rm C}$~=~8.39~$\pm$~0.05]{Asplund2005}, and the atomic and molecular line list compilation from \citet{AllendePrieto2018}, which includes CH, C$_{\rm 2}$, CN, and CO molecules. The assigned microturbulent velocity is determined by the star's effective temperature and surface gravity, further discussed in Section 2.3 of \citet{Adam2021}. \par

The computed 5D grid covers a large stellar atmospheric parameters space: 3500~$\leq$~T$_{\rm eff}$~$\leq$~6000~K (in steps of 250~K), 0.0~$\leq$~log(g)~$\leq$~5.0~cm~s$^{-2}$ (in steps of 0.5~cm~s$^{-2}$), -2.5~$\leq$ [M/H]~$\leq$~+0.5~dex (in steps of 0.5~dex), and the variation in [$\alpha$/M] is -0.25~$\leq$~[$\alpha$/M]~$\leq$~+0.75~dex, with steps of 0.25 dex. For each corresponding atmospheric parameter combination, the variation in the [C/M] element dimension  was originally from -0.25 to +0.25~dex, with steps of 0.25~dex, although we needed to extend it during this research up to -0.75~$\leq$~[C/M]~$\leq$~+0.5~dex. \par

It is also important to note that all metals in the synthetic grid are scaled by the same factor from the solar mixture (e.g.~[M/H]~=~0.2~=~[Fe/H]~=~[Li/H]), which implies by definition that [M/H]~=~[Fe/H], [$\alpha$/M]~=~[$\alpha$/Fe] and [C/M]~=~[C/Fe].

\subsubsection{Updated BOSZ synthetic stellar spectral library \citep{Bosz2024}} \label{method_BOSZ}

The new available set of BOSZ synthetic models is publicly available from MAST\footnote{\url{https://archive.stsci.edu/hlsp/bosz}} at the Space Telescope Science Institute, for eight different spectral resolutions from R~= 500 to 50000 and also at the highest sampling resolution of the synthesis, which varies between 200000 and 600000 depending on the main atmospheric parameters. These synthetic spectra cover a large wavelength domain (from 500 to 320000~$\AA$) . \par

In this work, we used the newly computed grid with LTE MARCS atmosphere models \citep{MARCS} and the solar reference abundances from \citet[][log~$\epsilon_{\rm C}$~=~8.39~$\pm$~0.05]{grevesse2007}, calculated for stars below T$_{\rm eff}$~$\leq$~8000~K. This MARCS grid has been previously implemented for the stellar parametrisation of the APOGEE 16th data release \citep{ahumada2020}, described in detail by \citep{Jonsson2020}.  In addition, this new MARCS grid has a larger parameter coverage than the one firstly presented in \citet{meszaros2012}. A comparison between the stars observed by the APOGEE survey and the synthetic grid coverage can be found in Fig.~1 in \citet{Bosz2024}. \par 

The new BOSZ synthetic stellar spectral library was computed with \texttt{SYNSPEC} \citep{Hubeny2017,  hubeny2021}, with important updates with respect to the molecular line list from \citet{AllendePrieto2018} by including the upgrades done by Kurucz\footnote{\url{http://kurucz.harvard.edu/}}. 
%including those from the ExoMOl project \citep{Tennyson2012, Tennyson2016}.
In particular, a significant improvement was the inclusion of the extensive list of CH from \citet{Masseron2014}. \par

Finally, the employed 5D BOSZ grid subset in the present study covers a wide atmospheric parameter and abundance range: 3500~$\leq$~T$_{\rm eff}$~$\leq$~6500~K (in steps of 250~K), 0.0~$\leq$~log(g)~$\leq$~5.0~cm~s$^{-2}$ (in steps of 0.5~cm~s$^{-2}$), -2.5~$\leq$ [M/H]~$\leq$~+0.75~dex (in steps of 0.25~dex), -0.25~$\leq$ [$\alpha$/M]~$\leq$~+0.5~dex (in steps of 0.25~dex), and -0.75~$\leq$ [C/M]~$\leq$~+0.5~dex (in steps of 0.25~dex). Similarly to the previous synthetic grid, [M/H] is what is typically called [Fe/H], scaling for all metals, which means that  [$\alpha$/M]~=~[$\alpha$/Fe] and [C/M]~=~[C/Fe].

\subsubsection{Differences between the adopted synthetic grids} \label{comparison_grids}

There are four main differences between the computation of stellar spectra in the BOSZ~\citep{Bosz2024} and \citet{Adam2021} synthetic grids that may account for discrepancies in carbon abundance determinations: \par 

\begin{enumerate}
    \item Within the effective temperature range explored in this work, the BOSZ grid employs MARCS stellar atmospheres from \citet{MARCS}, using spherical geometry for surface gravities log(g)~$<$~3.5 and plane-parallel geometry otherwise. In contrast, the \citet{Adam2021} models adopt ATLAS9 plane-parallel atmospheres from \citet{meszaros2012} across all temperatures and surface gravities. Comparisons between MARCS and ATLAS9 atmospheres are explored in \citet{meszaros2012, Bosz2024}.

    \item While both grids use atomic opacities and line lists detailed in \citet{AllendePrieto2018}, the BOSZ spectra include substantial updates to molecular opacities, incorporating 11 additional molecules detailed in Section 3.2 of \citet{Bosz2024}.

    \item The treatment of microturbulence differs between the two grids. The BOSZ grid provides spectra at fixed microturbulence values of 0, 1, 2, and 4 km/s. In contrast, the \citet{Adam2021} spectra use a variable microturbulence value determined by the star’s effective temperature and surface gravity, as described by equation 5 in \citet{Adam2021}. The absolute effect of microturbulence on line strengths can reach up to 1–2~$\AA$ in some cases. Further discussion of these effects can be found in Section 4.1 of \citet{Knowles2019} and references therein.

    \item Although both grids use \texttt{SYNSPEC} for spectral synthesis calculations, the adopted version in each is different. \citet{Adam2021} spectra were generated using the ASSET radiative transfer code \citep{Koesterke2009}, based on \texttt{SYNSPEC} routines and equations of state adopted from \citet{Hubeny2000, Hubeny2017}. BOSZ spectra are generated using the updated version of \texttt{SYNSPEC} given in \citet{hubeny2021}, that provides revisions to the equation of state, including improvements to the partition functions and number of molecular species considered in the calculations.  We direct readers to Section 3.3 of \citet{hubeny2021} for further details of the differences between \texttt{SYNSPEC} versions.
    
\end{enumerate}

We expect the biggest effect from these differences in models for the coolest stars with T$_{\rm eff}$~$\lesssim$~4000~K. Fortunately, as later described in Sect.~\ref{flags_and_errors}, we kept only stars with T$_{\rm eff}$~$\geq$~4000~K in our analysis following the suggestions by \citet{Xshooter_Param} based on the quality of the XSL parametrisation. Therefore, we did not find significant discrepancies among the models for the studied stellar sample. \par

%It is also worth noting that the parameter coverage of both grids is similar, with some differences in [alpha/M] and [M/H]. The \citet{Adam2021} grid extends to a higher value of 0.75~dex in [alpha/M], compared to 0.5~dex in the BOSZ grid. However, BOSZ offers a broader and more finely sampled range in [M/H], reaching up to +0.75~dex with 0.25~dex steps, compared to the +0.5~dex limit and 0.5~dex sampling of the \citet{Adam2021} grid. 

\subsection{Analysed CH bands: line selection} \label{method_lines}

Ideally, the most accurate way to determine [C/Fe] abundances in stars would be to analyse strong unblended atomic lines to avoid the dependence on the stellar parameters and the impact of the contribution of other chemical elements on the derived abundances. The atomic CI lines most commonly analysed in the literature are two unblended carbon lines at 5052.16 and 5380.34~$\AA$~\citep[e.g.][]{ecuvillon2004, DelgadoMena2010, DelgadoMena2021, nissen2014, amarsi2019}. These works showed a decrease of [C/Fe] with metallicity\footnote{ We will use [Fe/H] (fraction of iron over hydrogen) as the notation for the global stellar metallicity. The chemical abundance is expressed with respect to the Sun in a logarithmic scale: e.g.~[Fe/H]~=~log$\left[ \frac{N(Fe)}{N(H)} \right]_\star$ - log$ \left[ \frac{N(Fe)}{N(H)} \right]_\odot$, where N(X) is the density of the number of atoms for the element X.}, similar to that observed for the $\alpha$-elements (e.g. O, Mg, Si, S, Ca, Ti) evolution, suggesting an enrichment dominated by the core-collapse supernovae (Type~II~SNe). A slight increase in [C/Fe] in the -1.5~$\leq$~[Fe/H]~$\leq$~-1.0~dex range has also been observed in the Solar neighbourhood. The main problem of these atomic lines is that they become very weak towards lower metallicities ([Fe/H]~$\leq$~-1.5~dex) and cooler stars (T$_{\rm eff}$~$\lesssim$~5000~K), constraining the stellar selection. Furthermore, they could be affected by non-local thermodynamic equilibrium (NLTE) effects, showing differences with respect to 3D NLTE calculations compared to 1D LTE models \citep{amarsi2019}. Additionally, the forbidden [CI] line at 8727.13~$\AA$ has also been analysed for solar-type stars in the Galactic disc \citep{Andersson1994, Gustafsson1999, bensby2006}, observing a similar weak [C/Fe] decrease with increasing [Fe/H]. However, this line is very weak and blended (located in the left wing of a SiI line and blended by a FeI line), specially for stars with T$_{\rm eff}$~$\lesssim$~5700~K and high [Fe/H], being strongly required to have high spectral resolution (R~$\gtrsim$~60000) and signal-to-noise (S/N~$\gtrsim$~150).   \par

Secondly, several works have estimated [C/Fe] abundances in stars from some strong molecular CH bands in the G-band region around $\sim$4300~$\AA$ \citep[e.g.][]{carbon1987, alexeeva2015, zhao2016, SuarezAndres2017}, some of them compatible with those obtained from CI lines. CH strength is usually well correlated with the carbon abundance, showing a smooth behaviour. However, the reported trend in carbon with respect to [Fe/H] is flatter across all metallicity regimes. This CH region has the advantage of being detected in the entire metallicity range (at least down to [Fe/H]~$\gtrsim$~-2.5~dex), measurable even at low spectral resolution \citep[e.g.~R~$\sim$~1800 in LAMOST spectra,][]{Unni2022} and external galaxies \citep{Beverage2025}, and present minor influence of 3D NLTE effects on carbon abundance determinations~\citep{alexeeva2015}. In contrast, these lines are located in a crowded spectral region, where they could be blended or too strong to be properly measured \citep{Asplund2005, pavlenko2019}. A literature compilation of the observed [C/Fe]~vs.~[Fe/H] trend from both atomic and molecular carbon features will be later discussed in Sect.~\ref{results_literature}. \par

\begin{figure}
\centering
\includegraphics[height=73mm, width=0.45\textwidth]{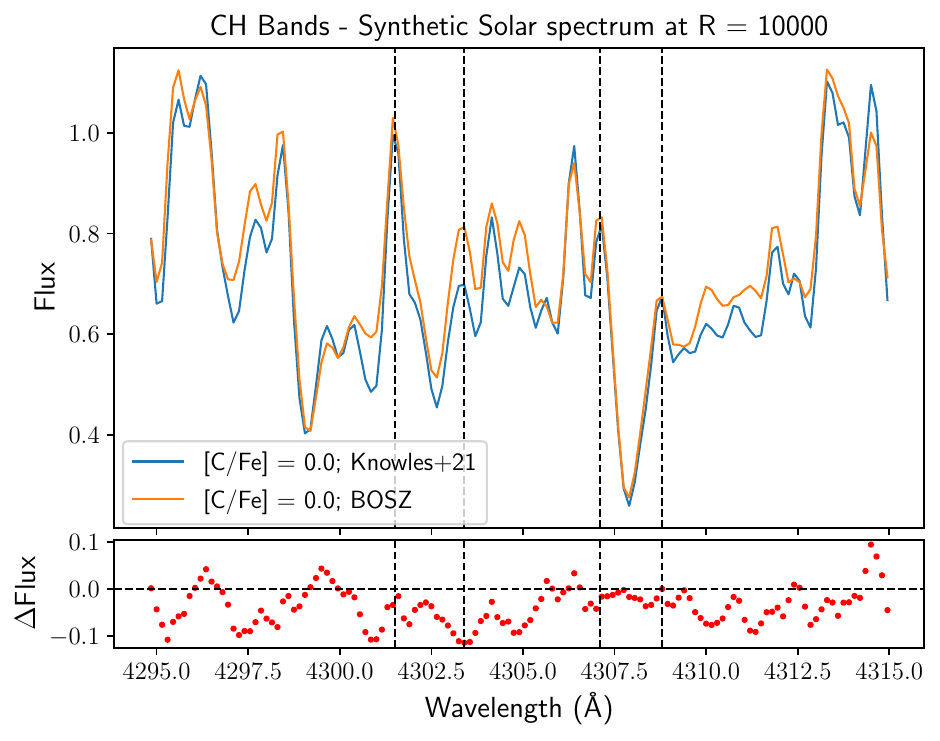}
\caption{Synthetic Solar spectrum in the CH bands region around $\sim$4300~$\AA$ at the X-shooter resolution (R~=~10000), from both employed reference grids: the one computed by \citet[][blue]{Adam2021}, and the updated BOSZ library (orange). The selected CH bands (see Table~\ref{table:lines}), where the [C/Fe] abundance is measured, are delimited by the black dashed vertical lines. The flux difference at each wavelength between both synthetic solar spectra is shown on the bottom panel.}
\label{Fig:grids_Sun}
\end{figure}

Regarding our analysis, the unblended CI lines 5052.16 and 5380.34~$\AA$ cannot be resolved at the X-shooter spectral resolution (R~$\sim$~10000), and therefore could not be properly measured. Consequently, the carbon abundance analysis was performed over strong molecular CH bands in the G-band region around $\sim$4300~$\AA$. Based on the previous work by \citet{SuarezAndres2017}, who selected ten clean spectral regions with strong CH features for solar-type stars at R~$\sim$~110000, we explored the suitability of these regions and re-adapted the analysis to the X-shooter resolution. Finally, the selected CH regions in the present study are shown in Table~\ref{table:lines} and illustrated in Fig.~\ref{Fig:grids_Sun}. These two CH bands have been tested in depth to ensure their measurability in any stellar type from the analysed X-shooter sample, i.e. independently from the chosen stellar atmospheric parameters (T$_{\rm eff}$, log(g), [Fe/H], [$\alpha$/Fe]).  \par

\begin{table}%[h]
\centering
\begin{tabular}{ccc}
\hline
\hline
\multicolumn{2}{c}{\textbf{CH bands (\AA):}}  \vspace{0.05cm} \\
%4279.2-4281.6 & 4301.5-4303.4 & 4307.1-4308.8 \\
4301.5-4303.4 & 4307.1-4308.8 \\
\hline
\hline
\end{tabular}
\vspace{0.08cm}
\caption{Carbon regions selected in the present analysis, adapted from \citet{SuarezAndres2017} to the X-shooter resolution (R~$\sim$~10000).}
\label{table:lines}
\end{table}

\begin{figure*}
\centering
\includegraphics[height=65mm, width=0.36\textwidth]{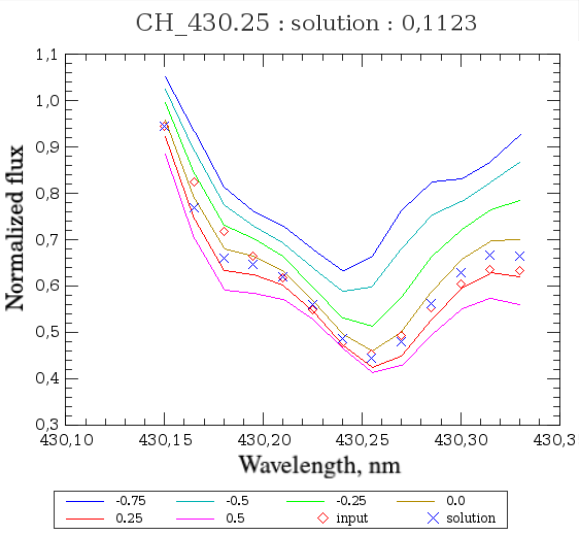}
\includegraphics[height=65mm, width=0.36\textwidth]{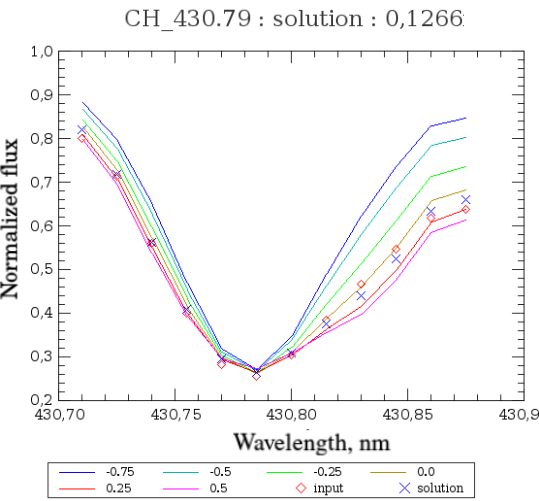}
\caption{Example of the fit carried out by the spectrum synthesis code GAUGUIN for the two CH analysed bands (showed in Table~\ref{table:lines}) in the Solar reflected spectrum of the Vesta asteroid, observed by the HARPS spectrograph~\citep{mayor2003}. It was convolved to the X-shooter spectral resolution (R~=~10000). The normalised observed spectrum is shown with red open diamonds, while the solution is indicated by blue crosses. The reference synthetic spectra grid \citep[from][]{Adam2021} is colour-coded according to the [C/Fe] abundance value.}
\label{Fig:GAUGUIN_fit_example}
\end{figure*}

Figure~\ref{Fig:grids_Sun} shows the synthetic Solar spectrum, convolved to the observed spectral resolution (R~=~10000), around the CH~4300~$\AA$ analysed region. The used reference synthetic spectra grids show some differences in the line profiles that may affect the final [C/Fe] abundance estimates (discussed in detail in Sect.~\ref{results}). It can be noticed how challenging it is to identify continuum in this crowded region, which will lead to a "pseudo-continuum” normalisation of the observed spectra when deriving the [C/Fe] abundance. \par

As described in detail in \citet[][see Sections~3.1 and 3.2 therein]{SantosPeral2020}, the GAUGUIN algorithm normalises the observed spectrum flux over a given wavelength interval centred on the analysed line, by comparing it to an interpolated synthetic one with the same atmospheric parameters. The most appropriate wavelength points of the residual (R~=~Synthetic/Observed) are selected by a $\sigma$-clipping iterative procedure. The residual trend is then fitted with a third-degree polynomial. The final normalised spectrum is obtained by dividing the original one by a linear function resulting from the fit of the residual \citep[see also the description in Sect.~6.3 in][]{alejandra2022_RVS}. We found that the abundance fit got worse when applying larger normalisation windows, resulting in a parameter dependence and increasing the band-to-band abundance dispersion. In this regard, in \citet{SantosPeral2020} we also studied the impact of the continuum definition in the precision and accuracy of the derived chemical abundances with GAUGUIN.  In this study, we followed the same optimised spectral normalisation procedure to guarantee the proper treatment of the data. \par

Figure~\ref{Fig:GAUGUIN_fit_example} illustrates the fit for the observed Solar spectrum at X-shooter resolution by the automated code GAUGUIN for the CH~4300~$\AA$ selected regions. In the case shown, the reference synthetic grid spectra corresponds to the one computed by \citet{Adam2021}, which was convolved and re-sampled accordingly. It can be appreciated the smooth correlated variations with the [C/Fe] abundance of the analysed CH bands. We obtained similar estimations of the carbon abundance from both bands, resulting in an average $\overline{\rm [C/Fe]}$~$\approx$~+0.12~dex. This bias implies a calibration from the Solar composition to gain accuracy. We verified the fit does not significantly improve or change the abundance measurement when we modified the abundance window interval, which highlights the robustness of the implemented method over these selected spectral regions. A similar analysis for the giant benchmark star Arcturus \citep[T$_{\rm eff}$~=~4286~K; log(g)~=~1.66~cm~s$^{\rm -2}$; {[}Fe/H{]}~=~-0.52~dex; ][]{RamirezAllendePrieto2011} can be found in Appendix~\ref{appendix_Arcturus}.

\subsection{Selection of the best analysed X-shooter sample} \label{flags_and_errors}

Based on the CH bands selection shown in Table~\ref{table:lines}, we only analysed the observed X-shooter spectra from the UVB arm (300-556~nm) in order to be able to derive the [C/Fe] stellar abundances. It consists of an original sample of 598 UVB spectra that corresponds to 504~stars. \par

We followed the same criteria described in \citet{SantosPeral2023} for the selection of the final XSL working sample (see Sect.~3.4 therein). In brief, we only kept the observed spectra with S/N~$\textgreater$~20~pixel$^{-1}$ (we discarded $\sim$~30 UVB spectra). Additionally, according to the error in the XSL parameter estimates at lower temperatures \citep{Xshooter_Param}, and the lack of valid synthetic spectra at T$_{\rm eff}$~$\textgreater$~6500~K due to sensitivity problems to line-broadening caused by stellar rotation and macroturbulence, we kept only those stars with 4000~$\leq$~T$_{\rm eff}$~$\leq$~6500~K to minimise the presence of inaccurate abundance values in the final catalogue. Moreover, none of the previously identified XSL carbon stars by \citet{Gonneau2016} are part of the studied stellar abundance sample. \par

%It is important to remark that we did not find carbon stars candidates in the obtained results hereafter. To this purpose, we followed the innovative analysis by \citet{lebzelter2018, abia2020, abia2022}, which includes the identification of a large number of new AGB carbon stars through their 2MASS photometry \citep{2MASS}, their \emph{Gaia} EDR3 astrometry \citep{GaiaEDR3, GaiaEDR3_Lindegren}, and their location in the \emph{Gaia}-2MASS photometric diagram. Therefore, we verified the absence of any possible AGB carbon stellar candidate in our final XSL sample. \par

%__________________________________________________________________

\section{Derived [C/Fe] abundances from the CH~band}\label{results}

For each individual X-shooter spectrum, the GAUGUIN algorithm yields separate [C/Fe] abundance measurements corresponding to each CH band listed in Table~\ref{table:lines}. To derive the final [C/Fe] abundance for each star, along with its associated internal uncertainties, we followed the methodology and quality criteria outlined in \citet{SantosPeral2023}. This approach accounts for the two independent abundance estimates obtained per spectrum —one for each CH band— as well as cases where multiple spectra are available for the same star. \par

In this regard, we estimated the sensitivity of the derived [C/Fe] abundances to the uncertainties on the input XSL stellar atmospheric parameters (T$_{\rm eff}$, log(g), [Fe/H]). As followed by \citet{SantosPeral2023}, we performed multiple Monte Carlo realisations of the atmospheric parameters associated to the input spectra. Table~\ref{table:error_parameters} illustrates the measured dispersion for different stellar types and metallicity regimes when using each synthetic grid separately. We adopt these dispersions as estimators of the uncertainties. The reported [C/Fe] abundance uncertainties are generally small, lower than $\sim$~0.1~dex, and the precision is similar for the two adopted reference grids in the study. \par

\begin{table*}
\centering
%\hspace*{-1.5cm}
%\vspace{+0.08cm}
\caption{Abundance uncertainty associated with the stellar atmospheric parameters.}
\begin{tabular}{cccc|ccc}
\hline
\hline
\\[\dimexpr-\normalbaselineskip+2pt]
 & Cool giant & Cool dwarf & Solar-type  & [Fe/H] & [Fe/H] & [Fe/H] \\
 & T$_{\rm eff}$ $\sim$ 4700 K & T$_{\rm eff}$ $\sim$ 4700 K & T$_{\rm eff}$ $\sim$ 5800 K  & ~$\textless$~-1.0~dex & ~-1.0~-~0.0~dex & ~$\textgreater$~0.0~dex \\
\\[\dimexpr-\normalbaselineskip+2pt]
\hline
\\[\dimexpr-\normalbaselineskip+2pt]
Knowles+21: $\Delta$[C/Fe]~(dex) & $\pm$~0.052 & $\pm$~0.049 & $\pm$~0.11 & $\pm$~0.057 & $\pm$~0.05 & $\pm$~0.033 \\
\\[\dimexpr-\normalbaselineskip+2pt]
BOSZ 2024 : $\Delta$[C/Fe]~(dex) & $\pm$~0.065 & $\pm$~0.066 & $\pm$~0.10 & $\pm$~0.097 & $\pm$~0.063 & $\pm$~0.056 \\
\\[\dimexpr-\normalbaselineskip+2pt]
\hline
\hline
\end{tabular}
\label{table:error_parameters}
\end{table*}

\begin{figure*}
\centering
\includegraphics[height=102mm, width=0.78\textwidth]{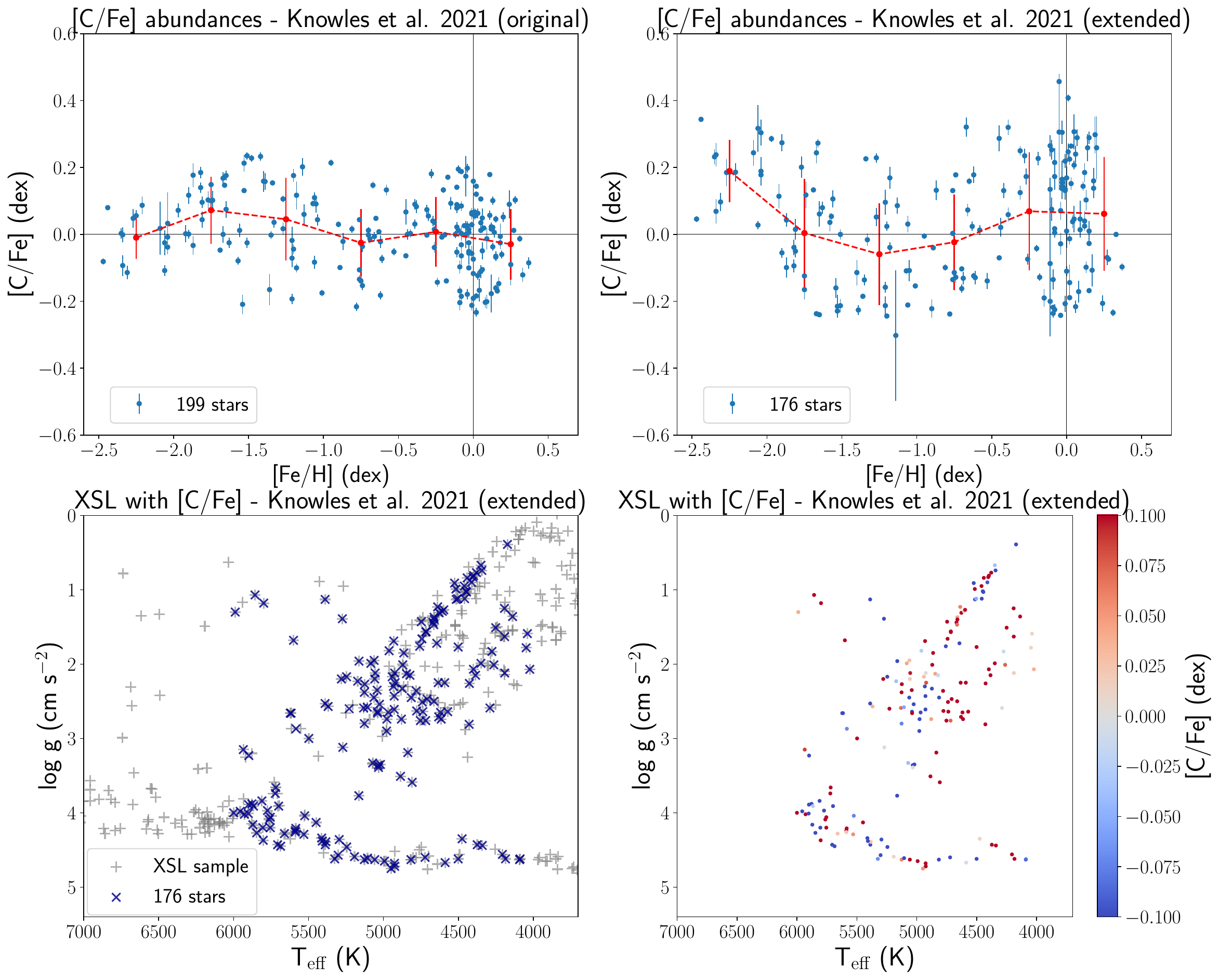}
\caption{\textit{Top row:} Stellar abundance ratios [C/Fe] vs. [Fe/H] of the X-shooter catalogue after applying the optimal methodology, with the original synthetic grid from \citet{Adam2021} (\textit{top-left}) and with the extended [C/Fe] dimension (\textit{top-right}), with the estimated internal uncertainties as vertical error bars. Each red point and error bar corresponds to the measured [C/Fe] average and scatter in metallicity bins of 0.5~dex. \textit{Bottom~row:}~HR diagrams of the final X-shooter stellar sample with reliable [C/Fe] abundance measurements from the extended grid: over the whole analysed X-shooter sample in this work (grey crosses, \textit{bottom-left}), and colour-coded by the carbon abundance estimate (grey crosses, \textit{bottom-right}).}
\label{Fig:CvsM_AdamGrids_HR}
\end{figure*}

We first performed an exhaustive analysis with the computed theoretical spectral library by \citet{Adam2021}, later complemented by the comparison with the more recent and updated BOSZ synthetic stellar spectral library \citep{Bosz2024}. Beforehand, for each used synthetic grid, we checked their accuracy by deriving [Mg/Fe] abundances from the X-shooter catalogue and comparing them to the well-validated measurements from \citet{SantosPeral2023}. This is illustrated in Appendix~\ref{appendix_Mg}.

\subsection{Results based on sMILES \citep{Adam2021}} \label{results_Adam}

Figure~\ref{Fig:CvsM_AdamGrids_HR} shows the measured [C/Fe] abundances as a function of the stellar metallicity [Fe/H] for the best analysed XSL sample, together with their estimated internal uncertainties, using the grid from \citet[][described in Sect.~\ref{method_Adam}]{Adam2021} before (top-left~panel) and after extending the [C/Fe] dimension (top-right~panel). It can be seen that, for both cases, the [C/Fe] abundances describe a dispersed ($\sigma$$_{\rm [C/Fe]}$~$\sim$~0.2-0.3~dex) flat trend around [C/Fe]~$\sim$~0.0~dex all along the covered metallicity range (-2.5~$\textless$~{[}Fe/H{]}~$\textless$~+0.5~dex). As mentioned in Sect.~\ref{method_lines}, previous works that analysed the CH molecular bands reported a flat trend with respect to metallicity (a complete comparison with the literature is discussed in Sect.~\ref{results_literature}). The main remarkable feature is the observed convex curve at low [C/Fe] abundances around [Fe/H]~$\sim$~-0.5~dex, which seems to be also present in some previous observational studies \citep[e.g. APOGEE DR16,][]{Jonsson2020}. With this analysis, we emphasise the importance of the grid dimension in the abundance space ([C/Fe]$_{i}$) since it clearly delimits the range of [C/Fe] estimates, larger and dispersed when we applied the extended synthetic grid, and also changes the stellar sample with high-quality [C/Fe] measurements (from 199 to 176 stars). For the last case, we also highlight the only presence of stars with [C/Fe]~$\textgreater$~0.0~dex in the very metal-poor regime ([Fe/H]~$\leq$~-2.0~dex). \par

\begin{figure*}
\centering
\includegraphics[height=85mm, width=0.7\textwidth]{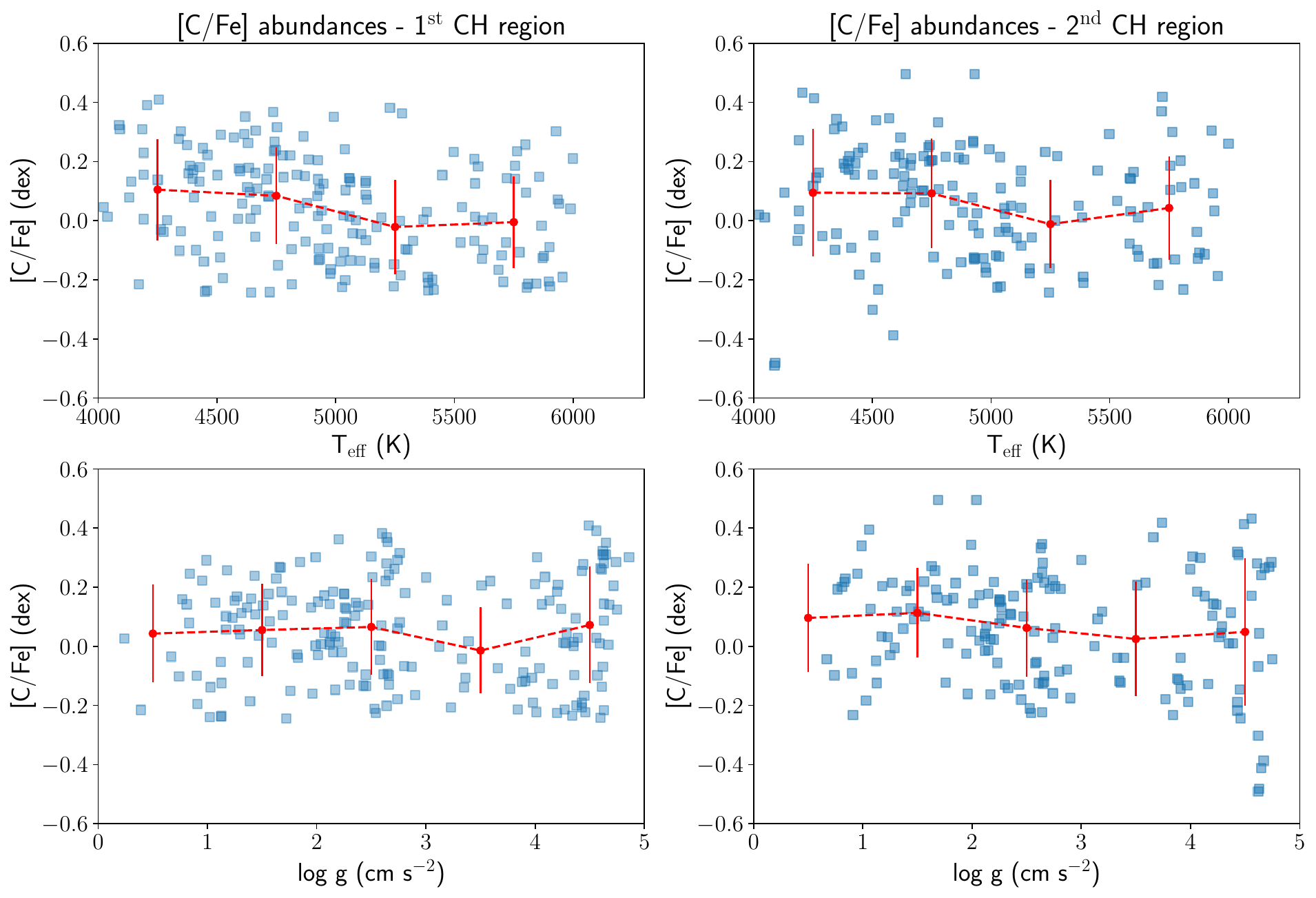}
\caption{Derived [C/Fe] abundance ratio from each analysed CH band separately (\textit{left and right columns}, shown in Table~\ref{table:lines}) as a function of the stellar effective temperature (\textit{top row}) and surface gravity (\textit{bottom row}). Each red point and error bar corresponds to the measured average and scatter in temperature and gravity bins of 500~K and log~1~cm~s$^{\rm -2}$, respectively.}
\label{Fig:C_AdamGrids_TeffLogg}
\end{figure*}

\begin{figure*}
\centering
\includegraphics[height=55mm, width=\textwidth]{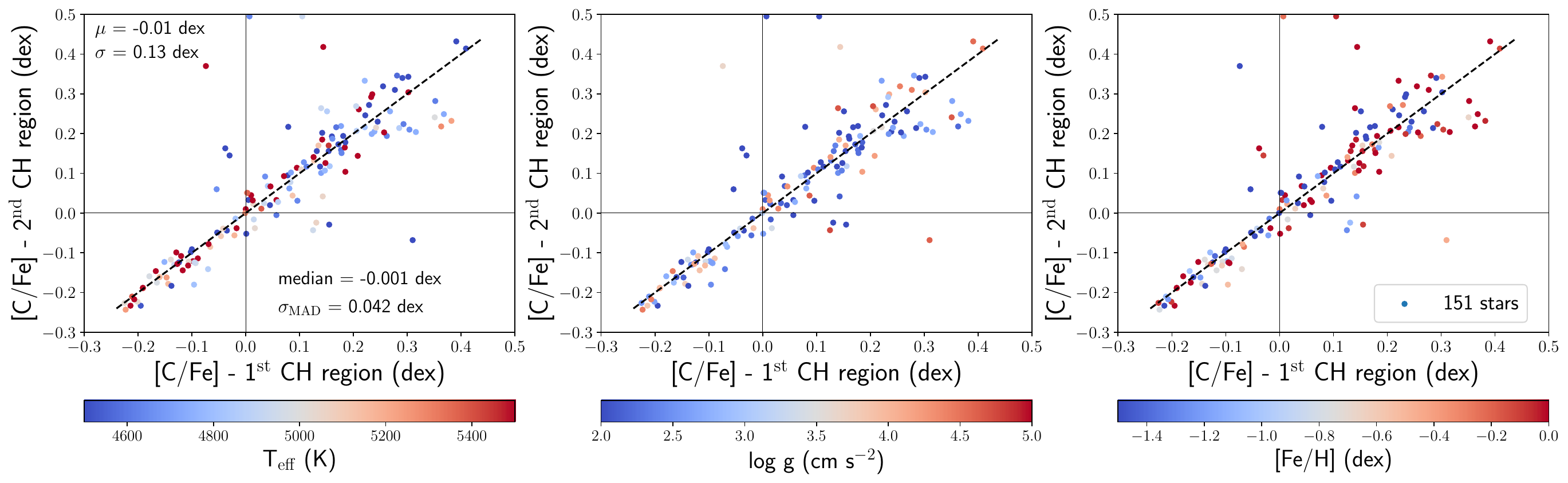}
\caption{Comparison between the derived [C/Fe] abundances from the individual analysed CH bands, colour-coded according to the effective temperature (\textit{left}), surface gravity (\textit{middle}), and metallicity (\textit{right}) of the star. The black dashed line reproduces the 1:1 relation. The mean ($\mu$), standard deviation ($\sigma$), median, and robust standard deviation (i.e. $\sim$~1.48 times the median absolute deviation, MAD) of the offsets are indicated in the first panel.}
\label{Fig:CH3vsCH2_TeffLoggFeH}
\end{figure*}

The covered parameters in the HR diagram for the final sample (using the extended grid) is also illustrated in Fig.~\ref{Fig:CvsM_AdamGrids_HR} (bottom row). We measured carbon abundances for different stellar types (67\% of the sample are giants), well distributed in the parameter space  (4000~$\lesssim$~T$_{\rm eff}$~$\lesssim$~6000~K~;~-2.5~$\lesssim$~[Fe/H]~$\lesssim$~+0.5~dex). Tentatively, the completeness of this sample in the atmospheric parameter and chemical space motivates the inclusion of carbon-enhancements in evolutionary stellar population synthesis models. \par

Additionally, it is worth noting that we do not observe any trend with the effective temperature or the surface gravity of the stars. Figure~\ref{Fig:C_AdamGrids_TeffLogg} shows a flat and dispersed trend of the derived carbon abundances from each individual CH band with respect to these atmospheric parameters. It is noteworthy that the reported convex curve at low [C/Fe] in Fig.~\ref{Fig:CvsM_AdamGrids_HR} is not appreciated in these diagrams, this feature is thus specific to the stellar metallicity. The unbiased parameter-dependence of the [C/Fe] abundance estimate constitutes a strong baseline for our research to provide a complete [C/Fe] abundance catalogue. \par

Furthermore, we evaluated possible abundance internal biases in our method by comparing the individual [C/Fe] estimates from each CH band, for the cases where they are both well-measured (151 stars). Figure~\ref{Fig:CH3vsCH2_TeffLoggFeH} shows the direct [C/Fe] band-to-band abundance comparison between the two analysed spectral regions, colour-coded with the star’s effective temperature, surface gravity and metallicity.  We generally observe a remarkable good agreement between the independent abundance estimates from each CH band, almost reproducing a perfect linear relation with identical [C/Fe] values (average offset of $\sim$~0.01~dex). This compatibility supports the robustness of the implemented method in deriving [C/Fe] abundances from individual CH molecular bands. \par

\begin{figure*}
\centering
\includegraphics[height=85mm, width=0.65\textwidth]{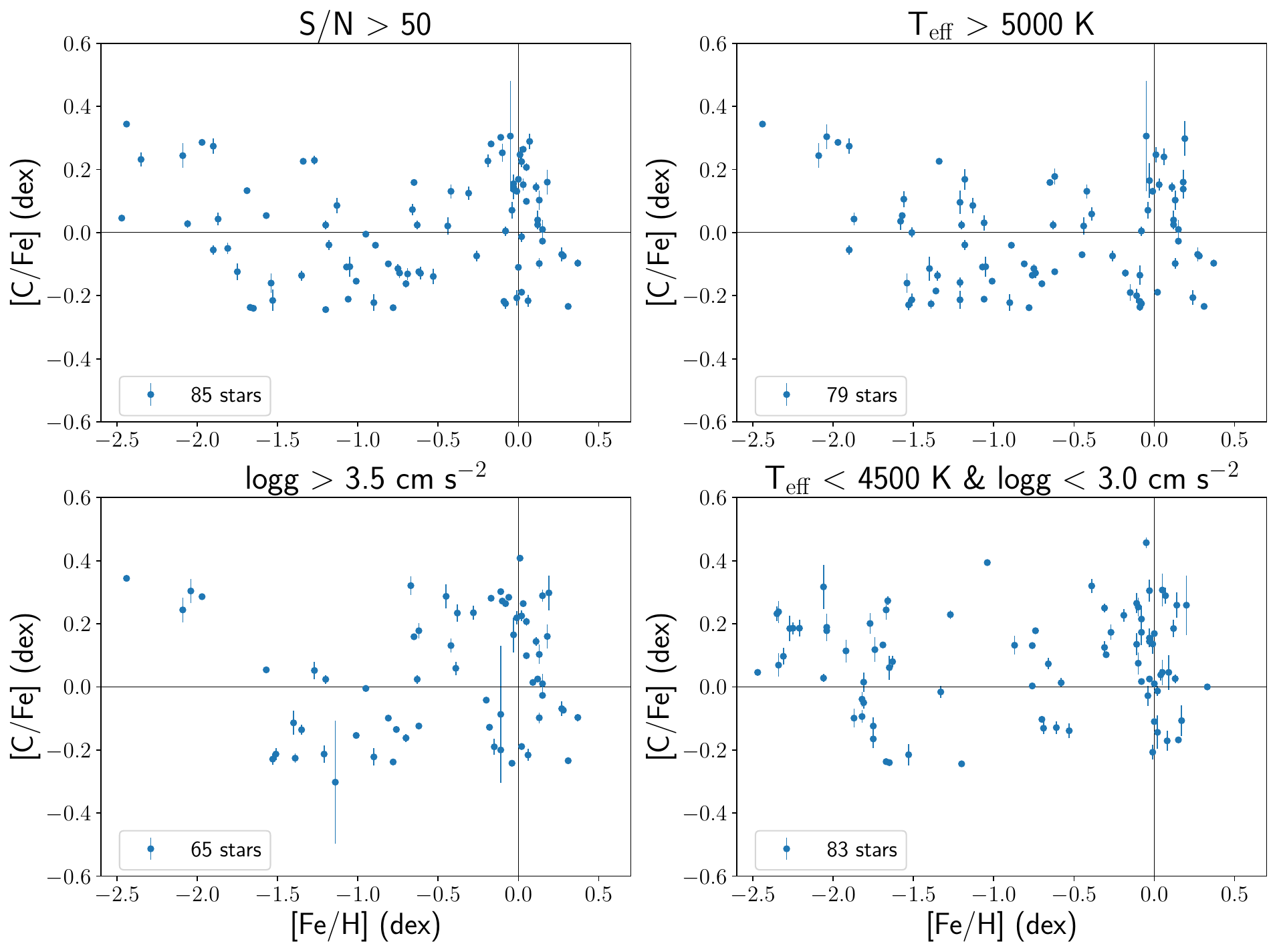}
\caption{[C/Fe] vs. [Fe/H] with the extended synthetic grid from \citet{Adam2021}, only those cases with high signal-to-noise (S/N~$\textgreater$~50, \textit{top-left}), hot stars (T$_{\rm eff}$~$\textgreater$~5000~K, \textit{top-right}), dwarfs (log(g)~$\textgreater$~3.5~cm~s$^{\rm -2}$, \textit{bottom-left}), and cool-giant stars (T$_{\rm eff}$~$\textless$~4500~K and log(g)~$\textless$~3.0~cm~s$^{\rm -2}$, \textit{bottom-right}).}
\label{Fig:CvsM_AdamGrid_HotDwarfs_SNR}
\end{figure*}

We observe slightly better fits in both CH analysed regions for hot dwarf stars (T$_{\rm eff}$~$\textgreater$~5000~K). This behaviour could be expected since the increased of the effective temperature reduces the presence of blended and molecular lines in the spectrum, and, according to the literature, the derived carbon abundance in these stars probably do not reflect the initial carbon composition with which the star was born because of the internal mixing in the atmospheres of giant stars \citep[e.g.][]{gallino1998, gratton2000}. Figure~\ref{Fig:CvsM_AdamGrid_HotDwarfs_SNR} illustrates the [C/Fe] abundance ratios, relative to the metallicity [Fe/H], for the most reliable stellar cases in the analysed X-shooter sample: those with S/N~$\textgreater$~50 (top-left panel), hot stars with T$_{\rm eff}$~$\textgreater$~5000~K (top-right panel), and dwarf stars exclusively (bottom-left panel). Additionally, we show those stars that could have passed through the first \textit{dredge-up} phase (T$_{\rm eff}$~$\textless$~4500~K and logg~$\textless$~3.0~cm~s$^{\rm -2}$, bottom-right panel). For each subsample, the reported stellar distribution is very similar with respect to the one previously described in Fig.~\ref{Fig:CvsM_AdamGrids_HR}. That is to say, a flat dispersed trend over the whole metallicity range, reproducing a convex curve at low [C/Fe] abundances around [Fe/H]~$\sim$~-0.5~dex, and only finding stars with [C/Fe]~$\textgreater$~0.0~dex in the very metal-poor regime. \par

The unbiased general trend shown in Fig.~\ref{Fig:CvsM_AdamGrid_HotDwarfs_SNR}, regardless of the selected stellar type or spectrum signal-to-noise ratio, does not allow us to distinguish any individual star in the final XSL sample as having a high-quality [C/Fe] abundance estimate. However, it is encouraging that we appear to be able to derive carbon abundances for both dwarfs and giants without distinction.

%__________________________________________________________________

\subsection{Results based on BOSZ \citep{Bosz2024}} \label{results_BOSZ}

Secondly, we implemented the updated BOSZ synthetic spectral library (described in detail in Sect.~\ref{method_BOSZ}) in GAUGUIN, as the new reference synthetic grid to derive the [C/Fe] abundances. We analysed the same CH molecular bands (Table~\ref{table:lines}) over the same selected XSL stellar sample illustrated before, in order to compare the results with the ones previously determined with the computed synthetic grid by \citet{Adam2021}. \par

\begin{figure*}
\centering
\includegraphics[height=65mm, width=0.36\textwidth]{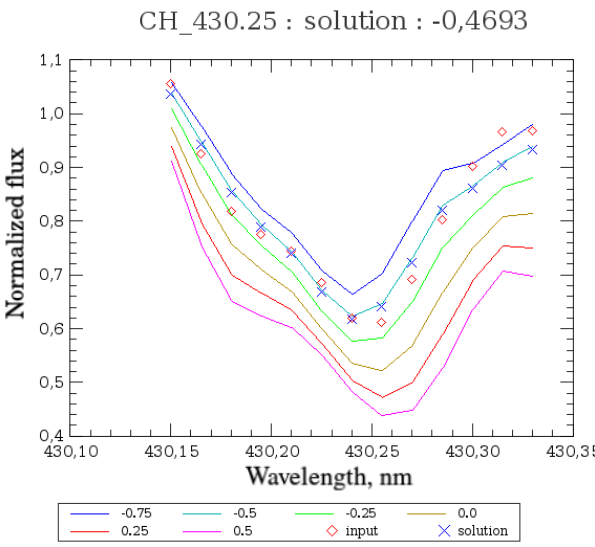}
\hspace{0.3cm}
\includegraphics[height=64mm, width=0.36\textwidth]{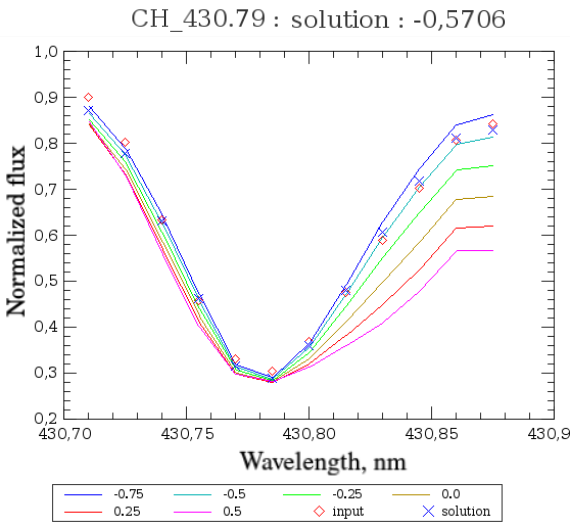}
\caption{Same as Fig.~\ref{Fig:GAUGUIN_fit_example}, but using the BOSZ synthetic spectra grid \citep[][]{Bosz2024} as a reference.}
\label{Fig:GAUGUIN_fit_example_BOSZ}
\end{figure*}

Figure~\ref{Fig:GAUGUIN_fit_example_BOSZ} illustrates the resulting fit for the observed Solar spectrum at the X-shooter resolution (see also for Arcturus in Appendix~\ref{appendix_Arcturus}). Similarly, as shown in previous Fig.~\ref{Fig:GAUGUIN_fit_example}, we observe a good fit and a smooth correlation of the [C/Fe] abundance in the analysed CH bands. However, the average [C/Fe] abundance of [C/Fe]~$\approx$~-0.52~dex is significantly different from the estimated with the \citet{Adam2021} grid ($\overline{\rm [C/Fe]}$~$\approx$~+0.12~dex), showing an averaged difference of $\arrowvert\Delta[\rm C/Fe]\arrowvert$~$\approx$~0.64~dex. This substantial disagreement could be explained by the differences in the lines profiles between the synthetic grids in the CH region (shown in Fig.~\ref{Fig:grids_Sun}) that may affect the [C/Fe] estimate procedure, although both grids were fully consistent in the [Mg/Fe] estimates (see Appendix~\ref{appendix_Mg}). \par

\begin{figure*}
\centering
\includegraphics[height=55mm, width=\textwidth]{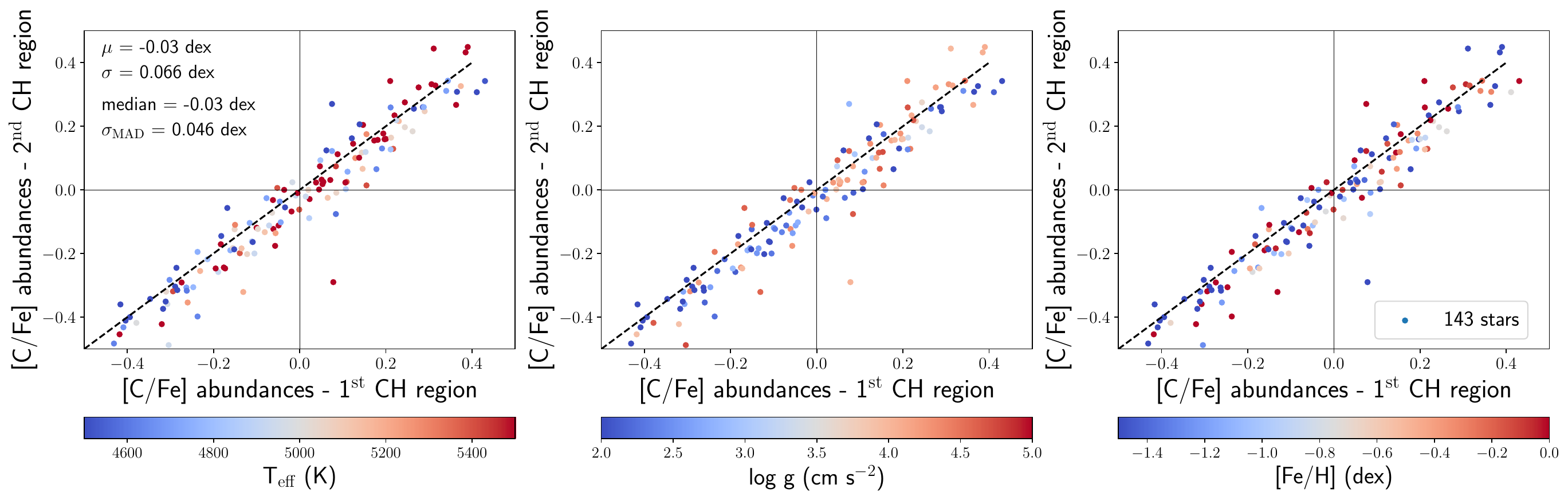}
\caption{Same as Fig.~\ref{Fig:CH3vsCH2_TeffLoggFeH}, but using the BOSZ spectral library for deriving the [C/Fe] stellar abundances.}
\label{Fig:CH3vsCH2_TeffLoggFeH_BOSZ}
\end{figure*}

In addition, we re-evaluated the comparison between the derived [C/Fe] abundances from each CH band. We also observe a very good agreement in Fig.~\ref{Fig:CH3vsCH2_TeffLoggFeH_BOSZ} (average offset of $\sim$0.03~dex), showing a strong compatibility. This indicates that the applied methodology remains robust in deriving [C/Fe] abundances from the CH molecular bands, as discussed in Fig.~\ref{Fig:CH3vsCH2_TeffLoggFeH} for the used grid from \citet{Adam2021}. We do not notice any trend with the stellar atmospheric parameters (i.e.~T$_{\rm eff}$, log g) either, and the final abundance sample is equally well distributed in the HR diagram parameter coverage. \par

\begin{figure*}
\centering
\includegraphics[height=50mm, width=\textwidth]{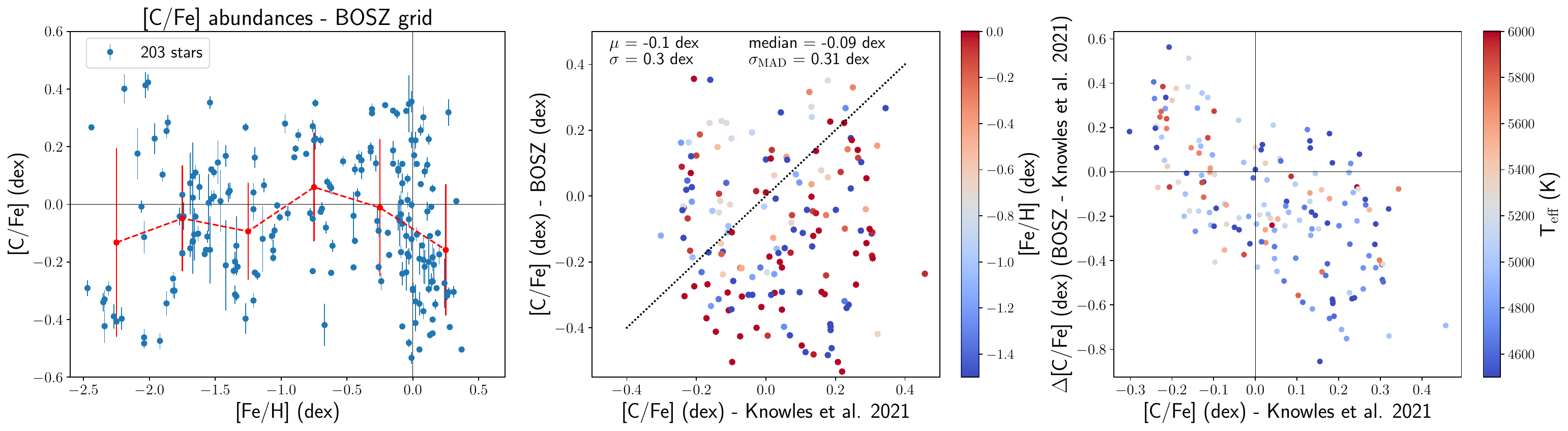}
\caption{\textit{Left:} Obtained abundance ratios [C/Fe] vs. [Fe/H] of the X-shooter catalogue with the BOSZ synthetic spectra grid \citep[][]{Bosz2024}. The red points are defined as in Fig.~\ref{Fig:CvsM_AdamGrids_HR}. \textit{Middle:} Direct comparison between the derived [C/Fe] abundances with each reference synthetic grid, colour-coded according to the star's metallicity. The black dashed line and the offset estimates are the same as those described in Fig.~\ref{Fig:CH3vsCH2_TeffLoggFeH}. \textit{Right:}~Difference in the measured [C/Fe] abundance between the used grids, colour-coded by the effective temperature, as a function of the [C/Fe] estimate previously determined with the grid from \citet{Adam2021}. }
\label{Fig:BOSZvsAdam_comparison}
\end{figure*}

Figure~\ref{Fig:BOSZvsAdam_comparison} shows the obtained [C/Fe] abundance ratio versus the metallicity [Fe/H] for the XSL sample (left panel), using the updated BOSZ synthetic spectral library \citep{Bosz2024}. We observe a similar, but more dispersed ($\sigma$$_{\rm [C/Fe]}$~$\sim$~0.4~dex), evolution of [C/Fe] to that depicted in Fig.~\ref{Fig:CvsM_AdamGrids_HR} with the \citet{Adam2021} grid. This behaviour was also found when we applied the same quality selections explained in Fig.~\ref{Fig:CvsM_AdamGrid_HotDwarfs_SNR}. In addition, we directly compared the [C/Fe] abundance results from each adopted reference synthetic grid (middle panel). It can be noticed that the analysed X-shooter stars are randomly distributed in a wide [C/Fe] abundance range (-0.4~dex~$\lesssim$~[C/Fe]~$\lesssim$~+0.4~dex), showing significant differences in absolute terms, up to $\arrowvert\Delta[\rm C/Fe]\arrowvert$~$\sim$~0.8~dex, implying a strong model dependence in the abundance estimate. Lastly, we also studied these differences as a function of the previous carbon estimates from \citet{Adam2021} (right panel). We observe a decreasing trend with [Fe/H], where most of the stars with low measured [C/Fe] abundances (e.g.~[C/Fe]~$\textless$~-0.2~dex) with the \citet{Adam2021} grid present the most enhanced [C/Fe] values when implementing the BOSZ grid from \citet{Bosz2024}. Surprisingly, although the overall [C/Fe]~vs.~[Fe/H] distribution looks similar, the individual stellar compositions show large variations with the synthetic grid. \par

%__________________________________________________________________

\subsection{Comparison with literature studies} \label{results_literature}

On the basis of the previous analysis, we were not able to resolve which set of stellar X-shooter [C/Fe] abundances are more accurate in accordance with an objective criterion. For this reason, we performed a detailed revision of carbon literature studies in order to draw more robust conclusions. \par

\begin{figure*}
\centering
\includegraphics[height=115mm, width=\textwidth]{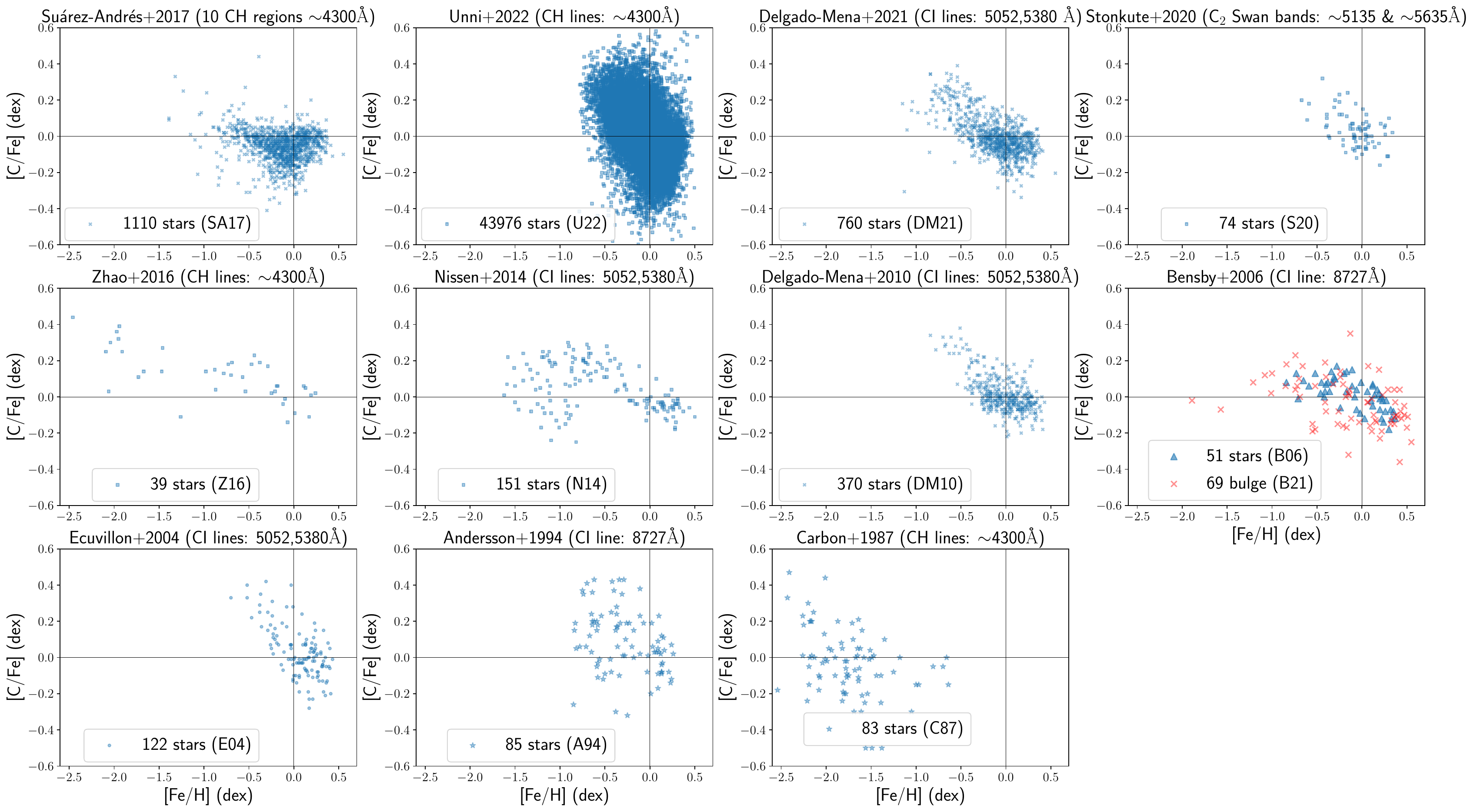}
\caption{[C/Fe] vs. [Fe/H] abundance ratios of previous literature studies with different carbon lines selection, as indicated on the top.}
\label{Fig:CvsM_Literature_comparison}
\end{figure*}

\begin{figure*}
\centering
\includegraphics[height=58mm, width=0.38\textwidth]{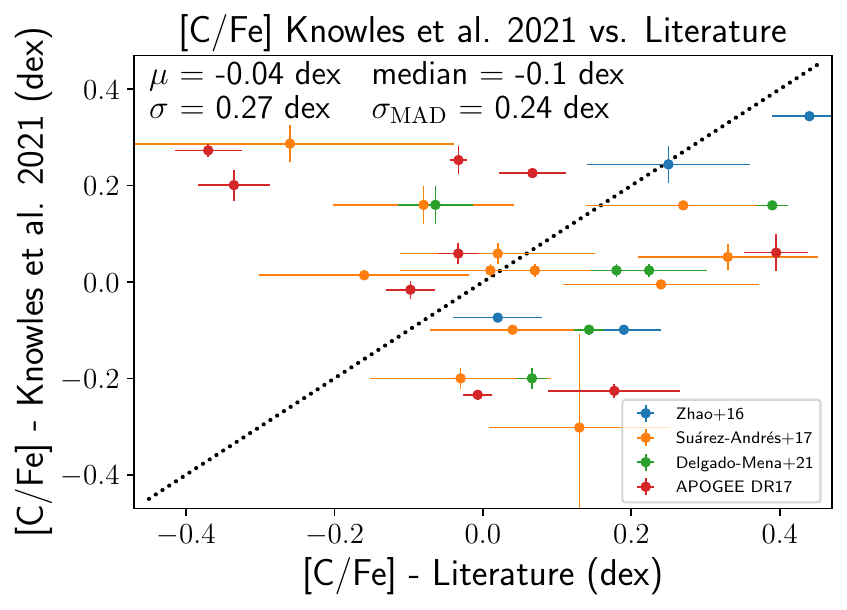}
\hspace{0.3cm}
\includegraphics[height=58mm, width=0.38\textwidth]{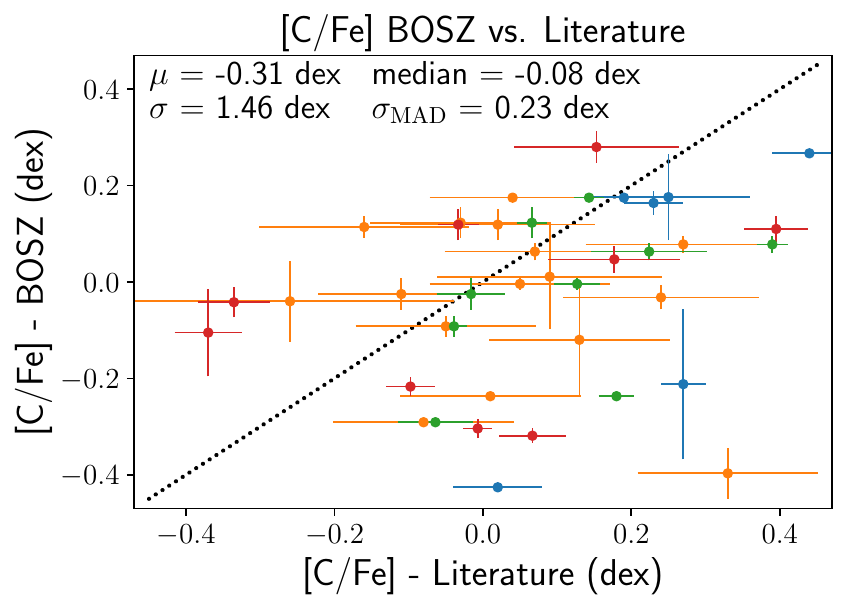}
\caption{Direct comparison between the derived
stellar abundance ratio [C/Fe] in this work, using the synthetic grid from \citet{Adam2021}~(\textit{left}) and the BOSZ library \citep[][\textit{right}]{Bosz2024}, and the abundance estimate for stars in common within the literature, with their respective reported uncertainties as error bars. The black dashed line are the same as in Fig.~\ref{Fig:CH3vsCH2_TeffLoggFeH}.}
\label{Fig:Direct_Literature_comparison}
\end{figure*}

Figure~\ref{Fig:CvsM_Literature_comparison} shows the [C/Fe]~abundance vs. metallicity~[Fe/H] plane from a large variety of published stellar catalogues in the literature. We remark the heterogeneity of results among the studies, even for those that analysed the same carbon lines. As mentioned before, these works selected only dwarf stars in their samples in order to avoid the potential misleading carbon measurements from the giants and AGB's atmospheres. Independently from the applied methodology, every work shows a large star-to-star dispersion in [C/Fe] at fixed metallicity, in agreement with what we observed in our study. In particular, the works that analysed atomic CI lines show a decrease of [C/Fe] with metallicity, similar to that observed for the $\alpha$-elements evolution. Moreover, we highlight the presence of a similar [C/Fe] convex shape around [Fe/H]~$\sim$~-0.5~dex in \citet{nissen2014} for the well-known CI lines (5052.16, 5380.34~$\AA$), which has also been illustrated by the APOGEE DR16 data \citep[see Fig.~12 in][]{Jonsson2020}. This feature, although slightly more diffuse, seems to be also present in \citet{SuarezAndres2017} for the CH~4300~$\AA$ region, which obtains a flatter carbon trend with respect to [Fe/H]. \par

Complementary, in Fig.~\ref{Fig:Direct_Literature_comparison} we directly compared our [C/Fe] abundance results (from both synthetic grids separately) for a subsample of stars in common within the literature \citep[APOGEE~DR17,][]{ApogeeDR17}. Previously, we verified the compatibility of the stellar parameters (T$_{\rm eff}$, log(g), [Fe/H]) among the different catalogues in order to be consistent, finding an excellent agreement, with no biases and almost identical values. However, regarding carbon abundances, Fig.~\ref{Fig:Direct_Literature_comparison} shows a very dispersed relation with respect to the literature, comparable for both adopted reference grids, presenting a similar median and $\sigma_{\rm MAD}$ estimates around -0.1 and 0.24~dex, respectively. Once again, the very close resemblance between the quality and accuracy of the different measured [C/Fe] abundances with the two employed synthetic spectral libraries do not allow us to prove which one would be more suitable for the carbon characterisation of the X-shooter Spectral Library.

%__________________________________________________________________

\subsection{Reported caveats} \label{results_caveats}

To summarize, the two employed 5D synthetic spectral libraries make possible a precise [C/Fe] abundance determination from CH molecular bands. The effect of the hydrogen atom seems to be negligible since [C/Fe] variations can be reproduced and measured at a given [Fe/H]. Every quality test satisfactorily pointed towards an achieved large unbiased [C/Fe] abundance catalogue, such as the quality of the line profile fit to derive the abundances (Figs.~\ref{Fig:GAUGUIN_fit_example}~and~\ref{Fig:GAUGUIN_fit_example_BOSZ}), the compatibility of the measured values between the analysed bands (Figs.~\ref{Fig:CH3vsCH2_TeffLoggFeH}~and~~\ref{Fig:CH3vsCH2_TeffLoggFeH_BOSZ}), or the lack of parameter dependence in the carbon estimate (Fig.~\ref{Fig:C_AdamGrids_TeffLogg}). In principle, these  catalogues seem ideal to shed new light on the carbon analysis problems and improving stellar population models in the near future. \par 

Nevertheless, we measured different and unpredictable [C/Fe] abundance results for the same star (up to $\arrowvert\Delta[\rm C/Fe]\arrowvert$~$\sim$~0.8~dex) with no apparent reason, just varying the reference synthetic grid (with equal [C/Fe] dimension) in the spectrum synthesis algorithm. They are well-proven theoretical grids that did not show such different abundance estimates for other chemical species (e.g [Mg/Fe], see Appendix~\ref{appendix_Mg}) with the same applied methodology. Therefore, the predominant model dependence in the [C/Fe] abundance measurements could arise from some hidden issues in the carbon modeling of the crowded and blended CH~4300~$\AA$ spectral region. Small differences in the synthetic models may induce a large disparity in the abundance estimate procedure (e.g.~observed spectrum normalisation, line fit), leading to completely disparate results. \par

\begin{figure}
\centering
\includegraphics[height=62mm, width=0.36\textwidth]{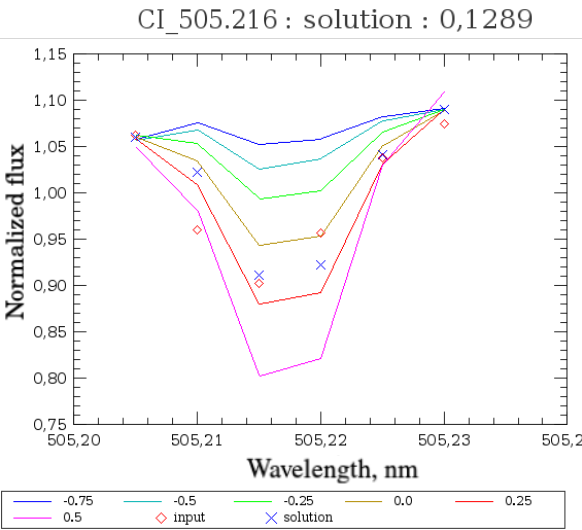}
\includegraphics[height=62mm, width=0.36\textwidth]{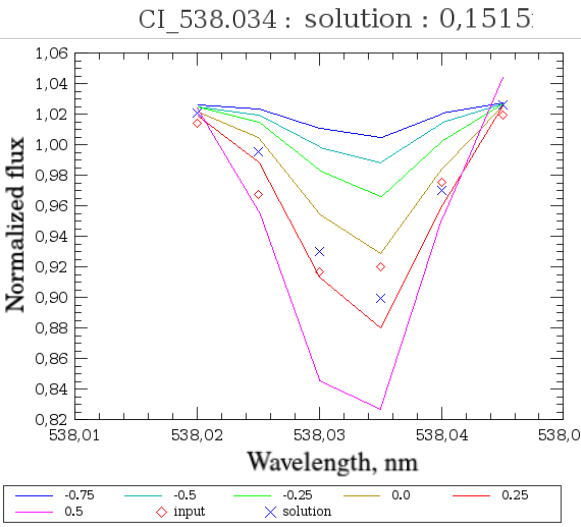}
\caption{Same as Fig.~\ref{Fig:GAUGUIN_fit_example}, but for the most used CI lines in the literature (5052.16, 5380.34~$\AA$) in the observed Solar spectrum at R~=~110000. The reference synthetic spectra grid is from \citet[][]{Adam2021}.}
\label{Fig:GAUGUIN_fit_example_CI}
\end{figure}

Complementarily, the medium spectral resolution of the X-shooter spectrograph (R~$\sim$~10000) could not be high enough for properly measuring carbon abundances. High spectral resolution (R~$\gtrsim$~50000) generally allows to disentangle blended lines and define a reliable continuum \citep{nissen2018}. On this basis, we analysed the typically studied unblended CI lines in the literature (5052.16, 5380.34~$\AA$) for the Solar spectrum at R~=~110000 \citep[same resolution as][who also analysed the CH regions]{SuarezAndres2017}. Figure~\ref{Fig:GAUGUIN_fit_example_CI} illustrates the obtained fit for each line separately. We observe that these atomic lines are less intense that the CH~4300~$\AA$~bands, also showing a smooth correlated variations with the [C/Fe] abundance.  We measured compatible abundance values from each CI line, $\overline{\rm [C/Fe]}$~$\approx$~+0.14~dex, which is similar to the one obtained at the X-shooter resolution from the CH regions ($\overline{\rm [C/Fe]}$~$\approx$~+0.12~dex, see Fig.~\ref{Fig:GAUGUIN_fit_example}). This compatibility of the derived [C/Fe] abundances for the Solar spectrum, with different resolutions and from different carbon lines, supports the robustness of the implemented synthetic spectrum algorithm in deriving [C/Fe] abundances. Consequently, the reported inconsistencies in our results might come from intrinsic issues in the carbon element modeling of the theoretical synthetic spectra.

With this analysis, we would like to note the potential error sources associated with the use of spectrum synthesis codes to derive stellar carbon abundances, which may lead to possible misunderstandings in the measured results for any stellar type.

%__________________________________________________________________

\section{Conclusions} \label{conclusions}

In this paper, we carried out a detailed spectroscopic analysis of the carbon abundance determination for the X-shooter Spectral Library (XSL), a large empirical stellar catalogue used as a benchmark for the development of stellar population models. We worked over the selected stellar sample by \citet{SantosPeral2023} for the previous $\alpha$-element (Mg, Ca) characterisation of the XSL, following the same abundance quality criteria. \par

We implemented the GAUGUIN spectrum synthesis code for deriving [C/Fe] abundances from individual CH molecular bands around the 4300~$\AA$ spectral region, since unblended atomic CI lines cannot be resolved at the XSL resolution (R~$\sim$~10000). The effect of the hydrogen atom seems to be negligible since [C/Fe] differences can be properly reproduced and measured from these bands at a given [Fe/H]. In order to evaluate the accuracy and the model dependence of the measured [C/Fe] abundances, we used two different computed sets of 5D (T$_{\rm eff}$, log(g), [M/H], [$\alpha$/Fe], and [C/Fe]) reference synthetic spectra grids: a theoretical library used for the computation of the semi-empirical MILES stellar spectra \citep[sMILES;][]{Adam2021}, and the updated BOSZ synthetic stellar spectral library \citep{Bosz2024}. We previously checked that they were fully consistent on deriving accurate [Mg/Fe] abundances. \par

From each adopted reference theoretical grid, we obtained a precise XSL [C/Fe] abundance catalogue of around 200 stars, with an excellent parameter coverage, showing no biases with the effective temperature or the surface gravity of the stars. Satisfactorily, we derived carbon abundances for both dwarfs and giants, with a similar high-quality [C/Fe] abundance estimate. In addition, we observe a remarkable good agreement between the individual [C/Fe] abundance measurements from each analysed CH band (with an average difference always lower than 0.03~dex), independently on the adopted synthetic grid. Moreover, we measured compatible [C/Fe] Solar abundance values from different spectral resolutions and atomic carbon lines, which supports the robustness of the implemented methodology, although a calibration is needed to reproduce the Solar composition. The final derived stellar abundance ratio [C/Fe], relative to the metallicity [Fe/H], always present a flat behaviour, dispersed ($\sigma$$_{\rm [C/Fe]}$~$\sim$~0.2-0.4~dex) around the zero [C/Fe] value for the whole covered metallicity range (-2.5~$\textless$~{[}Fe/H{]}~$\textless$~+0.5~dex). We do not observe a decrease trend of [C/Fe] with metallicity, as suggested by some works in the literature.  \par

We were able to provide a complete and unbiased [C/Fe] abundance catalogue with a well-tested robust method, which may constitute an important baseline research for including carbon-enhancements in the development of evolutionary stellar population synthesis models. However, we measured different, disparate and unpredictable [C/Fe] abundance values for the same analysed star (differences up to $\arrowvert\Delta[\rm C/Fe]\arrowvert$~$\sim$~0.8~dex) when we adopted one reference synthetic grid or the other in the applied spectrum synthesis algorithm. Consequently, there is a predominant model dependence in the [C/Fe] abundance measurements from the crowded and blended CH~4300~$\AA$ region. \par

Therefore, with the use of a spectral synthesis code to derive stellar carbon abundances in the CH~4300~$\AA$ band, we found that potential sources of error could be associated to unknown intrinsic constraints inside the carbon modeling of the theoretic synthetic spectra. This leads to inaccurate [C/Fe] abundance estimates for any stellar type, without significantly affecting the measured high-quality precision.

\begin{acknowledgements}
We thank the referee for his/her suggestions and discussions. The authors thank A. Recio-Blanco, P. de Laverny, C. Ordenovic and M. A. Álvarez for having developed the GAUGUIN method within the Gaia/GSP-spec context and having shared it with us. For any questions related to this algorithm, please contact directly A. Recio-Blanco and P. de Laverny. P.S.P and P.S.B acknowledges financial support by the Spanish Ministry of Science and Innovation through the research grant PID2019-107427GB-C31 and PID2022-138855NB-C3. P.S.P also acknowledges financial support by the European Union – NextGenerationEU under a Margarita Salas contract. A.V acknowledges support from grants PID2021-123313NA-I00 and PID2022-140869NB-I00 from the Spanish Ministry of Science and Innovation. This work has also been supported through the IAC project TRACES, which is partially supported through the state budget and the regional budget of the Consejer{\'{i}}a de Econom{\'{i}}a, Industria, Comercio y Conocimiento of the Canary Islands Autonomous Community. This work has made use of data from the European Space Agency (ESA) mission {\it Gaia} (\url{https://www.cosmos.esa.int/gaia}), processed by the {\it Gaia} Data Processing and Analysis Consortium (DPAC, \url{https://www.cosmos.esa.int/web/gaia/dpac/consortium}). Funding for the DPAC has been provided by national institutions, in particular the institutions participating in the {\it Gaia} Multilateral Agreement. Most of the calculations have been performed with the high-performance computing facility SIGAMM, hosted by the Observatorie de la Côte d'Azur (OCA).
\end{acknowledgements}

%-------------------------------------------------------------------

%%-----------------------------
%%   Bibliography
%%-----------------------------

%% The following lines are required when using BibTEX (strongly encouraged!):
\bibliographystyle{aa}  % A&A bibliography style file (aa.bst)
\bibliography{Santos-Peral2025} % your references in file: Yourfile.bib

\begin{appendix} 
\section{Arcturus} \label{appendix_Arcturus}

We reproduced the analysis performed for Figs.~\ref{Fig:GAUGUIN_fit_example} and \ref{Fig:GAUGUIN_fit_example_BOSZ}, considering the spectrum of a cool giant star instead of the solar one. We used Arcturus spectrum provided by \citet{Hinkle2000} as a benchmark, and adopted the atmospheric parameters from \citet{RamirezAllendePrieto2011}: T$_{\rm eff}$~=~4286~K; log(g)~=~1.66~cm~s$^{\rm -2}$; [Fe/H]~=~-0.52~dex; [C/Fe]~=~+0.43~$\pm$~0.07~dex. \par

\begin{figure*}
\centering
\includegraphics[height=68mm, width=0.38\textwidth]{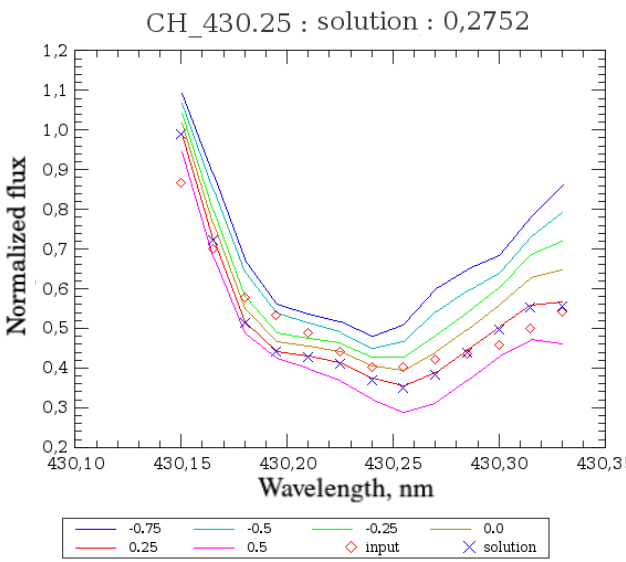}
\hspace{0.3cm}
\includegraphics[height=67mm, width=0.38\textwidth]{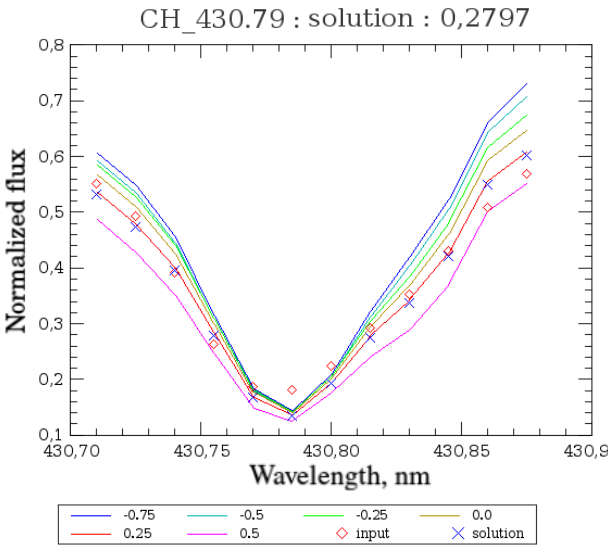}
\caption{Same fit as shown in Fig.~\ref{Fig:GAUGUIN_fit_example}, but for the observed Arcturus spectrum from \citet{Hinkle2000}.}
\label{Fig:GAUGUIN_fit_example_Arcturus}
\end{figure*}

\begin{figure*}
\centering
\includegraphics[height=68mm, width=0.38\textwidth]{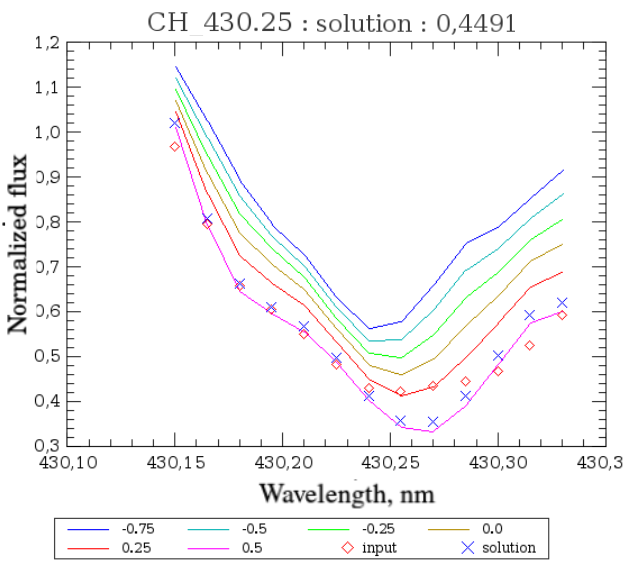}
\hspace{0.3cm}
\includegraphics[height=68mm, width=0.38\textwidth]{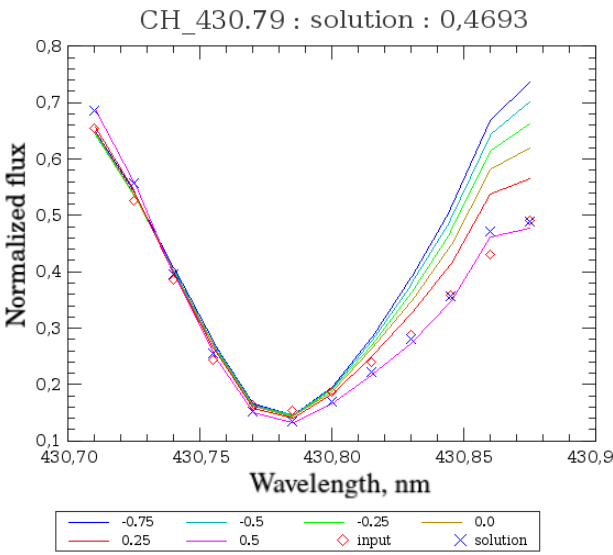}
\caption{Same as Fig.~\ref{Fig:GAUGUIN_fit_example_Arcturus}, but using the BOSZ synthetic spectra grid \citep[][]{Bosz2024} as a reference.}
\label{Fig:GAUGUIN_fit_example_Arcturus_BOSZ}
\end{figure*}

Figures~\ref{Fig:GAUGUIN_fit_example_Arcturus} and \ref{Fig:GAUGUIN_fit_example_Arcturus_BOSZ} show the obtained result from the same analysed CH bands (Table~\ref{table:lines}) with the \citet{Adam2021} synthetic grid and the BOSZ spectral library~\citep{Bosz2024}, respectively. Similar to the Solar case, we observe a good fit with a remarkable band-to-band abundance agreement from both grids, although different [C/Fe] estimates. We measured $\overline{\rm [C/Fe]}$~$\approx$~+0.28~dex with \citet{Adam2021} and $\overline{\rm [C/Fe]}$~$\approx$~+0.46~dex with BOSZ, being the latter compatible with the literature value. This leads to an averaged difference of $\arrowvert\Delta[\rm C/Fe]\arrowvert$~$\approx$~0.18~dex between both adopted grids. \par

Therefore, as discussed in the body of the paper, we obtained significant different [C/Fe] estimates depending on the adopted grid, but with a high-quality precision measurement. Unknown intrinsic issues in the carbon modeling may lead to inaccurate stellar carbon abundances.

\section{Preliminary validation analysis: [Mg/Fe] abundance determination} \label{appendix_Mg}

Previously to the carbon abundance analysis, we evaluated the quality and reliability of the used synthetic grids in order to avoid the potential propagation of uncertainties to the stellar abundance measurements. For this purpose, we firstly derived [Mg/Fe] abundances for the same X-shooter abundance catalogue published by \citet{SantosPeral2023}. \par

As described in detail in \citet{SantosPeral2020, SantosPeral2023}, the code GAUGUIN interpolates the reference synthetic spectra grid to the known input stellar parameters (T$_{\rm eff}$, log(g), [Fe/H], [$\alpha$/Fe]) of the observed star, generating a new specific-reference grid in the abundance space (e.g. [C/Fe]) to measure the abundance from the analysed spectral region. That is to say, the four input atmospheric parameters of each observed spectrum are fixed and required by GAUGUIN to derive individual chemical abundances. For this reason, it is crucial to verify if we can measure accurate $\alpha$-element abundances with the employed reference synthetic grids \citep[i.e.][]{Adam2021, Bosz2024} before extending the analysis to other chemical elements. \par

Therefore, we analysed the same strong well-known magnesium spectrum lines than in \citet{SantosPeral2023}, who used a completely different synthetic grid implemented for the AMBRE Project \citep{patrick2012} and the \emph{Gaia}-RVS analysis by the General Stellar Parametriser-spectroscopic (\emph{GSP-Spec}) module \citep{alejandra2022_RVS}. \par

\begin{figure*}
\centering
\includegraphics[height=100mm, width=\textwidth]{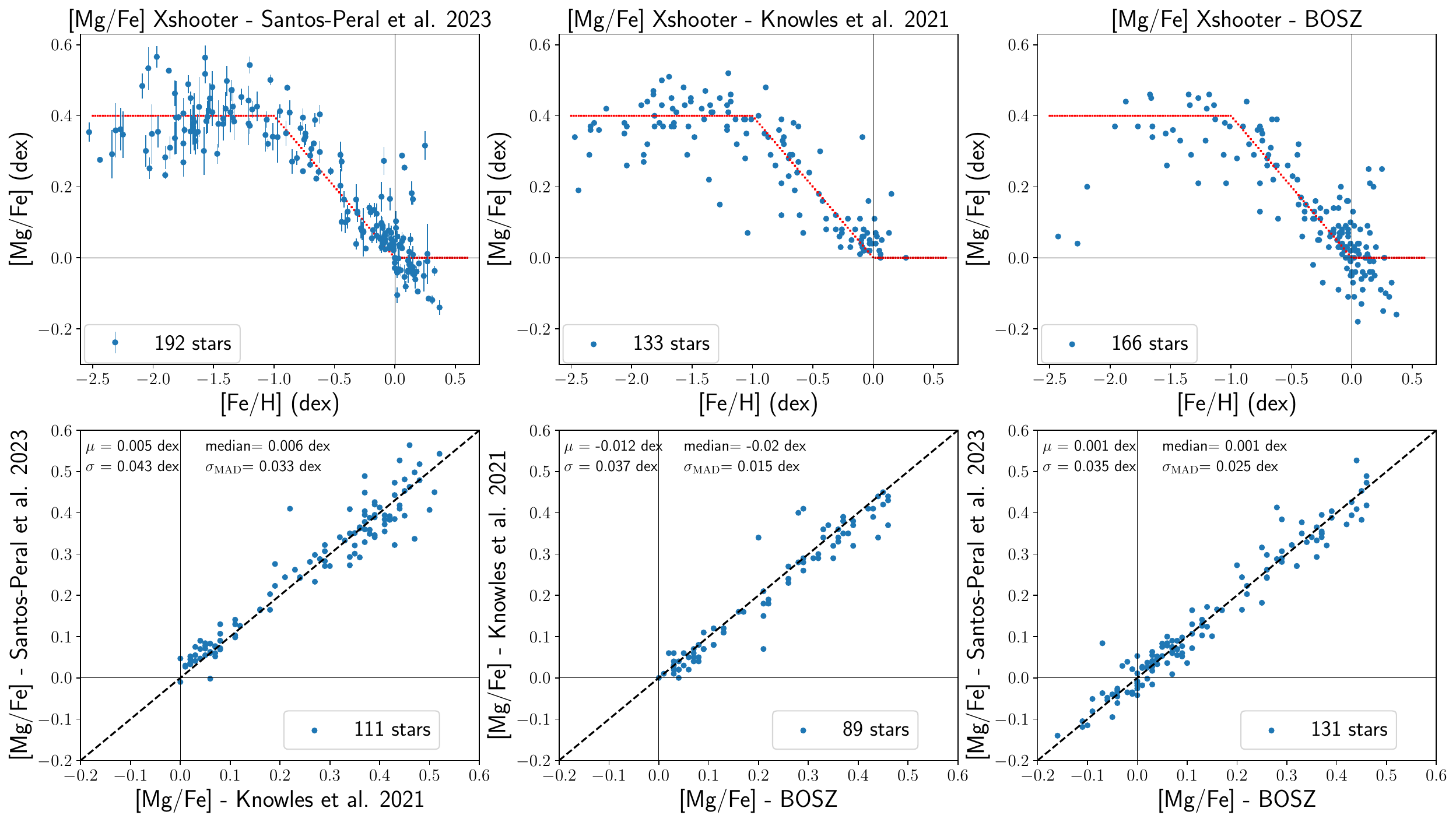}
\caption{Comparison between the derived [Mg/Fe] abundances of the X-shooter catalogue \citep{SantosPeral2023} and those measured when we used the synthetic grid by \citet{Adam2021} and the BOSZ library as the reference in the applied methodology. \textit{Top row:} stellar abundance ratios [Mg/Fe] vs. [Fe/H] from \citet{SantosPeral2023} (\textit{top-left}), with the grid by \citet{Adam2021} (\textit{top-middle}), and with BOSZ (\textit{top-right}). The red line reproduces an empirical [$\alpha$/Fe] approach for comparison purposes. \textit{Bottom row:} direct comparison of stars in common between the derived [Mg/Fe] abundances in \citet{SantosPeral2023} and those with \citet{Adam2021} (\textit{bottom-left}), between those with \citet{Adam2021} and with BOSZ (\textit{bottom-middle}), and finally between \citet{SantosPeral2023} and those with the BOSZ grid (\textit{bottom-right}). The black dashed line reproduces the 1:1 relation. The mean ($\mu$), standard deviation ($\sigma$), median, and robust standard deviation (i.e. $\sim$~1.48 times the median absolute deviation, MAD) of the offsets are also indicated.}
\label{Fig:Mg_comparison}
\end{figure*}

Figure~\ref{Fig:Mg_comparison} shows the comparison of the derived X-shooter [Mg/Fe] abundances among the three discussed synthetic spectra grids: the one used by \citet{SantosPeral2023}, and the two implemented in the present work: the computed one by \citet{Adam2021} and the updated BOSZ library \citep{Bosz2024}. We only plot stars with chemical abundance measurements that fulfilled the applied quality criteria. We observe a very good agreement between datasets, with small biases for each synthetic grid with respect to \citet{SantosPeral2023} results: averaged differences less than $\sim$~0.005~dex, with most of the stars with almost identical [Mg/Fe] values (bottom-left and bottom-right panels). The abundance estimates between the two grids used in the present work also present a remarkable agreement ($\arrowvert\overline{\Delta[\rm Mg/Fe]}\arrowvert$~$\sim$~0.02~dex, see bottom-middle panel) \par

In conclusion, the employed synthetic stellar spectral libraries in this study describe a robust behaviour in their theoretical atmospheric parameters (T$_{\rm eff}$, log(g), [Fe/H], [$\alpha$/Fe]), key for an accurate abundance determination.

\end{appendix}

\end{document}